\documentclass[%
 aip,
 amsmath,amssymb,
preprint,%
]{revtex4-2}

\usepackage{graphicx}
\usepackage{dcolumn}
\usepackage{bm}
\usepackage[mathlines]{lineno}

\usepackage[utf8]{inputenc}
\usepackage[T1]{fontenc}
\usepackage{mathptmx}
\usepackage{etoolbox}
\usepackage{float}
\usepackage{xcolor}
\usepackage[caption=false]{subfig}

\makeatletter
\def\@email#1#2{%
 \endgroup
 \patchcmd{\titleblock@produce}
  {\frontmatter@RRAPformat}
  {\frontmatter@RRAPformat{\produce@RRAP{*#1\href{mailto:#2}{#2}}}\frontmatter@RRAPformat}
  {}{}
}%
\makeatother
\begin{document}

\preprint{AIP/123-QED}

\title[The \textbf{$\Upsilon$} \textit{indicator} for Early Warning]{
Dynamical Stability Indicator based on Autoregressive Moving-Average Models:  Critical Transitions and the Atlantic Meridional Overturning Circulation}
\author{Marie Rodal}
 \altaffiliation{marie.rodal@fu-berlin.de}
\affiliation{ 
FB Mathematik und Informatik, Freie Universität Berlin, Arnimallee 6, 14195 Berlin, Germany
}%

\author{Sebastian Krumscheid}
\affiliation{Karlsruhe Institute of Technology, 76131 Karlsruhe, Germany}
\author{Gaurav Madan}
\author{Joseph Henry LaCasce}

\author{Nikki Vercauteren}

 \altaffiliation{nikki.vercauteren@geo.uio.no}
\affiliation{
 Section for Meteorology and Oceanography, Department of Geosciences, University of Oslo, Blindernveien 31, Kristine Bonnevies hus, 0371 Oslo, Norway
}
\date{\today}

\newcommand{\change}[1]{{\color{blue}#1}}

\newcommand{\changeSK}[1]{{\color{purple}#1}}

\begin{abstract}
A statistical indicator for dynamic stability known as the $\Upsilon$ indicator is used to gauge the stability and hence detect approaching tipping points of simulation data from a reduced 5-box model of the North-Atlantic Meridional Overturning Circulation (AMOC) exposed to a time dependent hosing function. The hosing function simulates the influx of fresh water due to the melting of the Greenland ice sheet and increased precipitation in the North Atlantic. The $\Upsilon$ indicator is designed to detect changes in the memory properties of the dynamics, and is based on fitting ARMA (auto-regressive moving-average) models in a sliding window approach to time series data. An increase in memory properties is interpreted as a sign of dynamical instability.  
The performance of the indicator is tested on time series subject to different types of tipping, namely bifurcation-induced, noise-induced and rate-induced tipping. The numerical analysis show that the indicator indeed responds to the different types of induced instabilities. Finally, the indicator is applied to two AMOC time series from a full complexity Earth systems model (CESM2). Compared with the doubling CO$_2$ scenario, the quadrupling CO$_2$ scenario results in stronger dynamical instability of the AMOC during its weakening phase. 
\end{abstract}

\maketitle

\begin{quotation}
A statistical indicator for dynamic stability is applied to simulation data from an ocean circulation model. The indicator assesses the stability of the time series data and gives indication of approaching tipping points. Three different types of tipping, defined by their causing mechanism, are explored. In addition, the indicator's reaction to the application of colored, as opposed to white, noise is assessed. Finally, the indicator is compared to other statistical early warning indicators. 
\end{quotation}

\section{\label{sec:level1}Introduction}
Tipping points, or critical transitions, are sudden, drastic changes in a system resulting from initial small perturbations. The study of tipping points is of particular interest to climate scientists and ecologists, as several theoretical studies highlight such tipping for an assortment of climatic and ecological systems, and observations also indicate that abrupt changes are, indeed, common in nature \cite{lenton_environmental_2013}. \\
\citet{Ashwin2012} classified tipping points according to the causing mechanism, yielding three classes of tipping points. Bifurcation-induced tipping, or B-tipping, occurs when a steady change in a parameter past a threshold induces a sudden qualitative change in the system's behaviour. Noise-induced tipping, or N-tipping, occurs when short-timescale internal variability causes the system to transition between different co-existing attracting states. Finally, rate-induced tipping, or R-tipping, occurs when the system fails to track a continuously changing attractor and hence abruptly leaves the attractor. \\
Of these three, rate-induced tipping is certainly the least studied, however as demonstrated by \citet{Scheffer2008}, \citet{Wieczorek2011} and more recently \citet{O'Keeffe2019}, it is an important tipping mechanism that cannot be explained through classical bifurcation theory. Indeed, when the system is unable to track a continuously available quasi-stable state due to the system parameters changing too quickly, it might shift to another available equilibrium state without crossing a bifurcation boundary. There are a few methods available for estimating what exactly "too quickly" means, see \citet{Wieczorek2014}, \citet{Ashwin2017}, \citet{Feudel2011}  and \citet{O'Keeffe2019}, but they depend strongly on the time-dependent parameter function; in particular its asymptotic properties. Finding generalizable methods for determining the rate of the parameter drift that induces tipping, will be of great interest going forward. Another issue of great practical importance is the question of how to obtain early warnings for such tipping points, in particular if classical methods for stability analysis also remain valid in the regime of rapid parameter changes.\\
\citet{RitchieSieber2016} showed that for rate-induced tipping, the  most commonly used early-warning indicators, namely increase in variance and increase in autocorrelation, occur not when the equilibrium drift is fastest but with a delay. This suggests that these indicators might not be able to detect tipping before it has already occurred, although their analysis does give indication that the theory behind these indicators, the so-called "critical slowing down", may still hold for rate-induced tipping. \\
In this paper, we study an indicator for dynamic stability, from now on referred to as the \textbf{$\Upsilon$} \textit{indicator}, initially proposed by \citet{Faranda2015}. The \textbf{$\Upsilon$} \textit{indicator} uses auto-regressive moving-average or ARMA(p,q) models to estimate how close a system is to an equilibrium. It is based on the observation that the dynamics of an observable arising from a potentially complex system very close to a stable equilibrium will appear like a random walk with a tendency to be attracted to a well-defined equilibrium. When discretized, such dynamics can be well represented by an ARMA(1,0) process. When approaching a transition, however, the system may experience a critical slowing down and diverging memory properties. The trajectory of the observable hence experiences new timescales, which can be detected even with a limited dataset through an increase in the necessary memory lags of fitted ARMA(p,q) models \cite{Faranda2014}. The \textbf{$\Upsilon$} indicator thus defines a distance from the limiting random walk-like behaviour as a way to assess the dynamical stability properties of an observable. The indicator was applied to atmospheric boundary layer data by \citet{Nevo2017} and \citet{Kaiser2020} and to atmospheric circulation data by \citet{Faranda2016}. They successfully demonstrated the indicator's ability to both gauge the stability of a time series and detect tipping points. However, the indicator requires some additional testing, in particular concerning its performance for 
rate-induced tipping, which thus far has not been explored. It should be noted that several different early warning indicators based on ARMA models have been proposed. In fact, in \citet{Faranda2014} the authors propose the sum of the p and q orders of the model, as well as the sum of the model coefficients as potential indicators. The sum of the order parameters then gives an estimate for the memory lag of the process, while the sum of the model coefficients gives the persistence of this memory lag. \\
To further test the indicator, we have chosen the global oceanic 3-box model studied by \citet{Alkhayuon2019}, which in turn is based upon the 5-box model of \citet{Wood2019}. The model represents a simplified Atlantic Meridional Overturning Circulation (AMOC), which transports warm surface water from the tropics to North America and Europe, resulting in a milder climate in these regions than what would otherwise be expected. Since the current is density driven, a large influx of freshwater due to the melting of land ice or increased precipitation in the North Atlantic, would be expected to result in a reduction in the AMOC flow strength. The question of whether the AMOC could undergo a sudden transition from a high flow strength state (the "on" state) to a state with weak or no overturning (the "off" state), is still debated. 
The latest assessment report of the International Panel for Climate Change (IPCC AR6) concludes that the AMOC strength will very likely decline in the future, but states with medium confidence that an abrupt collapse will not occur in the next century \cite{IPCCAR6}. Simple box models, like the one presented in this paper, show bi-stability, while more realistic models like the global atmosphere-ocean general circulation models (AOGCMs) are largely mono-stable, implying that they do not exhibit the abrupt transition to an "off"-state so characteristic of the simpler models. However, there is limited evidence that the more complex models may be too stable (\citet{Weijer2019}, \citet{HofmanRahmsdorf2009} and \citet{Liu2017}), in particular that they mis-represent the direction of AMOC-induced freshwater transport across the southern boundary of the Atlantic (\citet{Liu2017}, \citet{Huisman2009}, \citet{Liu2009}, \citet{hawkins2011}). \citet{Liu2017} demonstrated that by introducing a flux-correction term into the National Center for Atmospheric Research (NCAR) Community Climate System Model version 3 (CCSM3), they could make the formerly mono-stable system bi-stable. \\
In addition, it has been suggested that paleoclimate data is consistent with abrupt changes in the surface temperature in the North Atlantic region in the past, as might be expected with a collapse of the AMOC. 
\citet{boers_observation-based_2021} applied a statistical early warning indicator on Earth System Model (ESM) outputs, and found significant early-warning signals in eight independent AMOC indices. This was interpreted as a sign that the AMOC is not only a bistable system, but one approaching a critical transition. 
\\\\
Previously, the potential collapse of the AMOC has largely been attributed to the crossing of a bifurcation boundary in the bi-stable system. However, more recent analysis, see in particular \citet{LohmanDitlevsen2021}, demonstrate the possibility of tipping before the bifurcation boundary is reached through the mechanism of rate-induced tipping. In addition, \citet{LohmanDitlevsen2021} demonstrate that due to the chaotic nature of complex systems a well-defined critical rate, i.e., the rate of parameter change at which the system tips, cannot be obtained, which in turn severely limits our ability to predict the long-term behavior of the system. They conclude that due to this added level of uncertainty, it is possible that the safe operating space with regard to future emissions of CO$_2$ might be smaller than previously thought. This suggests that proper evaluation of the probability of rate-induced tipping in the different tipping elements of the Earth System is of utmost importance in assessing the likelihood of dramatic future changes.
\\
Regardless of whether the AMOC in actuality is bi-stable or mono-stable, the reduced 5-box model of \citet{Alkhayuon2019} is the perfect test case for the \textbf{$\Upsilon$} \textit{indicator} as it exhibits both bifurcation-induced and rate-induced tipping, provided a time dependent hosing function is applied. The hosing function represents the influx of fresh water into the ocean due to increased precipitation and melting of land and sea ice in the North Atlantic region. \citet{Alkhayuon2019} provide an extensive analysis of the tipping mechanisms present in the model. 
Armed with such a well studied theoretical model, we will be able to systematically study the indicator's ability to not only detect bifurcation-induced and noise-induced, but also rate-induced tipping. 
We will additionally assess the indicator's ability to deal with colored noise, something that is known to cause issues for other early warning indicators, like the increase in variance and auto-correlation \cite{boers_observation-based_2021}.  

In reality, the ocean system has many more degrees of freedom than those included in the box models, and ultimately a mixture of different processes is likely to trigger tipping, if occurring. The Coupled Model Intercomparison Project (CMIP6), with the Community Earth System Model (CESM2)\cite{Danabasoglu_2019}, provides an alternative AMOC model with many more degrees of freedom. Two scenarios where the atmospheric CO$_2$ concentration is abruptly increased will be considered, providing monthly outputs of geographical density differences on which the \textbf{$\Upsilon$} \textit{indicator} will be applied. In these model scenarios, the abrupt change in CO$_2$ is followed by a response of the Earth system, and after 2-3 decades, freshwater eventually circulates in the sub-polar gyre \citep{Madan2022}. This response hence offers similarities with the hosing experiments done in the box models. While the two scenarios are insufficient to assess the potential bistability of the AMOC, the \textbf{$\Upsilon$} indicator will be used to assess the dynamical stability of the AMOC during its weakening phase.
\\

\section{The $\Upsilon$-indicator for early-warning signals}
In what follows, we will briefly outline the method used to determine the stability of the time series data. Further details can be found in \citet{Faranda2015}, \citet{Faranda2016},  \citet{Nevo2017} and \citet{Kaiser2020}\\
The method relies on an accurate representation of a complex dynamical system close to a metastable state by a random walk-like behavior with a tendency to be attracted to the metastable state. Changes in the system's stability are then characterized as statistically significant deviations from that local behavior, indicating that the system currently does not reside close to a metastable state. Indeed, the local dynamics of a continuous-time random dynamical system (i.e., a stochastic differential equation) near a metastable state come close to the dynamics of a stochastic spring (i.e., an Ornstein--Uhlenbeck process), whose discrete-time observations are well approximated by an ARMA (1,0) process.
Here, ARMA denotes the space of autoregressive moving-average models, with the numbers in parentheses denoting the order of the model. A time series $x(t), \;t \in \mathbf{Z}$, is an ARMA(p,q) process if it is stationary and can be written as 
\begin{equation}
\label{eq:ARMA_process}
    x(t) =  \nu + \sum_{i=1}^p\phi_ix_{t-i} + \sum_{j=1}^{q}\theta_jw_{t-j} + w_t
\end{equation}
with constant $\nu$, coefficients $\phi_i$, $\theta_j$ and $\{w_t\}$ being white noise with positive variance $\sigma^2$ (see \citet{brockwell_introduction_2002} for an introductory text). In addition,  
constraints are imposed on the coefficients $\phi_i$ and $\theta_j$ to ensure that the process in \eqref{eq:ARMA_process} is stationary and  
satisfies the 
invertibility condition. 
Intuitively, the variables $p$ and $q$ say something about the memory lag of the process, while the prefactors $\phi_i$ and $\theta_j$ relate to the persistence of said memory lag. One expects that the higher the values for $q$ and $p$, the longer the system, once perturbed from its equilibrium state, would need to return to equilibrium. It is this intuitive notion that the statistical indicator denoted $\Upsilon$ takes advantage of. Indeed, when approaching a critical transition the response of the system to perturbations can become increasingly long (referred to as a critical slow down), and this translates into diverging memory properties of the statistical signal. Hence, an ARMA(p,q) model will require higher orders to incorporate the memory effects. By fitting the model \eqref{eq:ARMA_process} repeatedly to a time series data set for varying values of $p$ and $q$, one can, through application of an appropriate information criterion, obtain the values of $p$ and $q$ that best represent the time series data. For this purpose, we choose the Bayesian information criterion, BIC:
\begin{equation}
\label{eq:BIC}
\text{BIC} = -2 \ln L(\hat{\beta})+\ln(\tau)( p + q + 1 ) 
\end{equation}
where $\hat{\beta}$ denotes the maximum likelihood estimator of $\beta = (\nu, \phi_1,\dots,\phi_p,\theta_1,\dots,\theta_q)$, which is obtained by maximising the likelihood function $L$ associated with the ARMA(p,q) model \eqref{eq:ARMA_process} for a given time series; see \citet{brockwell_introduction_2002} for details. The best fitting ARMA(p,q) model is then determined as the one that minimizes the BIC.
The second term in equation \eqref{eq:BIC} punishes complex models with high $p$ and $q$ values, and is the reason why we prefer to use the BIC over other criteria, such as the perhaps more familiar 
Akaike Information Criterion. Here, $\tau$ denotes the number of discrete points in the time series to which the ARMA model is fitted. We refer to $\tau$ as the \textit{window length}. \\
Finally, the stability indicator is defined as
\begin{equation}
\label{eq:Upsilon}
\Upsilon(p,q;\tau) = 1 - \exp\left(\frac{-\left|\text{BIC}(\bar{p},\bar{q})-\text{BIC}(p,q)\right|}{\tau}\right)
\end{equation}
where $\bar{p}$ and $\bar{q}$ indicate the order of what we refer to as the theorized \textit{base model}. This is the ARMA(p,q) model, characterized by a specific value of $q=\bar{q}$ and $p=\bar{p}$, to which the chosen best fit is compared.
The $\Upsilon$-indicator takes on values between 0 and 1, where lower values imply a higher degree of stability. The intuition behind using the difference in BIC values between the chosen "best" model and a base model is that this quantity assesses just how much better the model with the lower BIC value approximates the fitted data compared to the other. The significance threshold for deviations in the BIC values between an ARMA(p,q) and the base model, simply denoted as $|\Delta$BIC$|$, is $|\Delta$BIC$|> 2$. The differences in BIC values 
can be directly related to the Bayes Factor, see \citet{Preacher2012}, which is another way of quantifying the likelihood of one model over another. \\
For the data sets analysed by \citet{Faranda2015}, it was determined that the appropriate base model is the ARMA(1,0) model, i.e., $\bar{p}=1$ and $\bar{q}=0$, which can be viewed as a time discretized Langevin process. In later work by \citet{Nevo2017} and \citet{Kaiser2020} the authors continued to rely on ARMA(1,0) as the base model. While \citet{Faranda2015} used a statistical argument to justify the choice of the base model, \citet{Nevo2017} and \citet{Kaiser2020} argued, as already noted above,  that the dynamics near a stable state can be approximated as that of a stochastic spring, further strengthening the case for ARMA(1,0) as the general choice of base model. However, due to the additional well-posedness constraints on the autoregressive and moving-average coefficients $\phi_i$ and $\theta_j$ in \eqref{eq:ARMA_process}, depending on the treatment of constraints by the fitting routine one can have cases where the BIC value of the ARMA(1,0) process is smaller than the corresponding value for the chosen ARMA(p,q) model. In these cases the ARMA(1,0) process is rejected as the best fit, despite having the lowest BIC value, due to violating the stationarity or invertibility conditions required for a numerically well behaved fit. Thus, in this scenario it becomes unclear how to determine the 'distance' between the states. To overcome this issue we have chosen to modify the $\Upsilon$ indicator to allow for a second base state, namely the ARMA(0,0) model. This model is just white noise, possibly with a drift, and is guaranteed to satisfy all the auxiliary conditions for the obvious reasons that there are no coefficients available to violate them. We consider ARMA(0,0) as a special case of ARMA(1,0) in which $\phi_1=0$. The use of the ARMA(1,0) process as a base model was partly justified by the image of a particle trapped in a potential well, where a restoring force keeps the particle oscillating around the equilibrium. The justification for including ARMA(0,0) as a potential base model follows a similar argument, except that in this case the noise amplitude is too low compared to the width of the potential well to feel the restoring force. To use both base models, we first introduce
\begin{equation}
    \Delta \text{BIC}_0(p,q) := \text{BIC}(0,0) - \text{BIC}(p,q) 
\end{equation}
and
\begin{equation}    
    \Delta \text{BIC}_1(p,q) := \text{BIC}(1,0) - \text{BIC}(p,q)
\end{equation}
With this, the modified $\Upsilon$-Indicator for the extended base model class can be written as
\begin{equation}
\label{eq:Upsilon_modefied}
    \Upsilon(p,q;\tau) = 1 - \exp\left(\frac{-\text{min}\left\{|\Delta \text{BIC}_0(p,q)|, |\Delta \text{BIC}_1(p,q)|\right\}}{\tau}\right)
\end{equation}
In addition, it must be specified that in the cases where the constrained fitting failed for the ARMA(1,0) model so that $\Delta \text{BIC}_1(p,q)$ may be negative, $\Delta \text{BIC}_0(p,q)$ is automatically chosen in practise. For obvious reasons, there cannot be a case where $\Delta \text{BIC}_0(p,q)$ is itself negative.
\\\\
Furthermore, following \citet{Faranda2014}, we define the \textit{order}, $\mathcal{O}$, and \textit{persistence}, $\mathcal{R}$, of an ARMA$(p,q)$ process as 
\begin{eqnarray}
\label{eq:order}
    \mathcal{O} &=& p + q\;,\\
\label{eq:persistence}
    \mathcal{R}&=& \sum_{i=1}^p |\phi_i| + \sum_{i=1}^q |\theta_j|\;,
\end{eqnarray}
where $\phi_i$ and $\theta_j$ denote the autoregressive and moving-average coefficients, respectively. While the order relates to the memory lag of the process, the persistence relates to the \textit{persistence} of said memory lag, hence the name. When approaching a tipping point, one would expect one out of two things to happen: either both the persistence and the order increase significantly, due to the increased memory of the process, or the order remains constant, and the persistence approaches the value of the order $\mathcal{O}$, indicating a loss of stationarity. According to \citet{Faranda2014}, the latter alternative corresponds to a case in which the potential landscape of the system does not change considerably when approaching the transition.\\
This observation strengthens the case for the modified $\Upsilon$ indicator in contrast to
excluding windows of the time series where $\Delta\text{BIC}_1(p,q)$ is negative, as these periods are indicative of an instability resulting from the loss of stationarity of the ARMA(1,0) process.
\\\\
To apply the method to a time series data set, one first has to ensure stationarity of the data. This can be done in two ways, depending on the nature of the time series. In some cases, it is sufficient to split the time series into small enough intervals, so that within each interval the time series is approximately stationary. To check for stationarity one runs a Kwiatkowski–Phillips–Schmidt–Shin (KPSS) tests on the intervals. This way, one also obtains an upper bound on the length of the intervals; see \citet{Kaiser2020}. The other option is to not assume stationarity from the outset, and instead allow for application of a differencing routine to the separate intervals, achieving stationarity that way. In that case, a KPSS test is run on each interval, and if the interval is found to not be stationary, differencing is applied. This process is then repeated until stationarity is achieved. The KPSS test is to be preferred over the unit root test due to the danger of over-differencing (\citet{Hyndman2008}). As we wish to study rate induced tipping phenomena, which yields highly non-stationary time series even for very small interval lengths, the latter method is to be preferred. By this choice we go from an ARMA to an ARIMA model, in which the \textit{I} stands for "integrated" in reference to the differencing routine used to ensure the stationarity of the time series.\\
Provided one can select sufficiently long 
time series intervals where the process is approximately stationary, one can fit ARMA(p,q) models to available observations during these intervals, and through the $\Upsilon$ indicator obtain an estimate for how close any given interval is to an equilibrium state. To determine the best fit, we use the auto.arima function found in the FORECAST R package, setting BIC as the information criterion used for model selection. Since we will not assume stationarity of the time series, auto.arima first determines the correct differencing order before continuing with the fitting procedure; the details of said procedure can be found in \citet{Hyndman2008}.\\
It is clear that the method is strongly dependent upon the size of the intervals, which we will refer to as the window length, $\tau$. This is not only due to the inclusion of the $1/\tau$ factor in the exponential, but also due to the inherent $\tau$-dependence of  BIC$(p,q)$ and BIC$(1,0)$. In fact, the rationale for including the $1/\tau$ factor in the definition of \textbf{$\Upsilon$} is to attempt to remove or reduce this dependence. From equation (2) one might conclude that the correct scaling would be $1/\ln(\tau)$, as opposed to $1/\tau$. However, we do not only want to remove the dependence on $\tau$, but also include the significance threshold for $\Delta$BIC, such that the \textbf{$\Upsilon$} value of any point where $\Delta$BIC is below 2 is suppressed relative to other points. 
\\\\
\begin{figure}
\includegraphics[width = 0.8\linewidth]{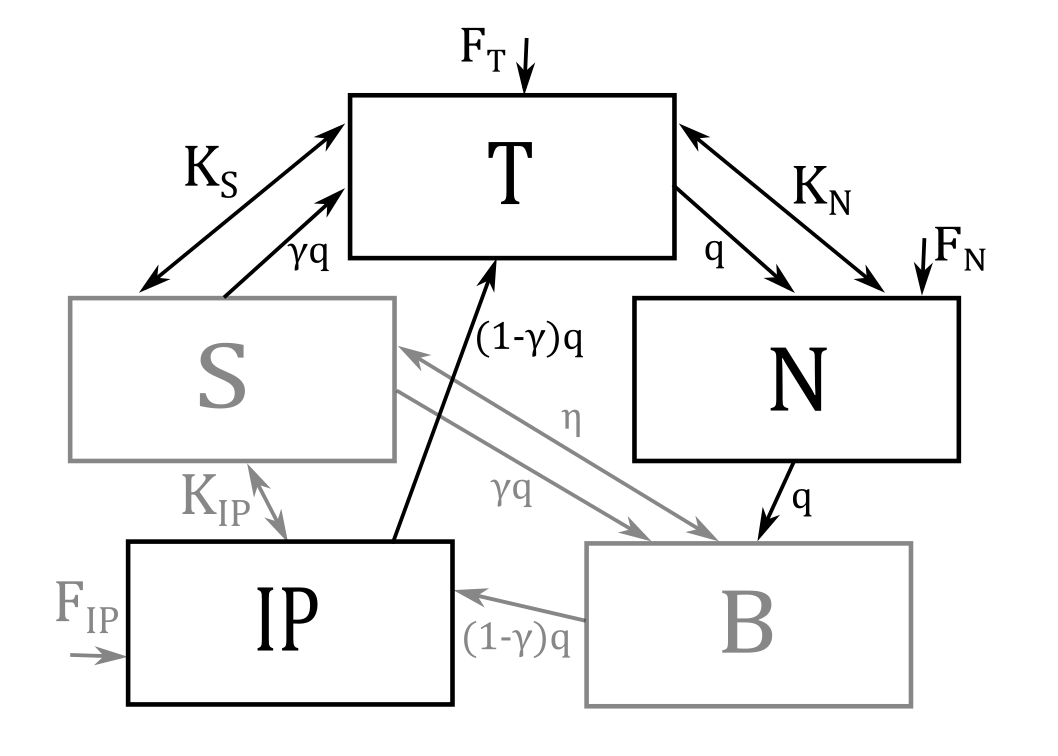}
\caption{Sketch of the 5-box model for the Atlantic Meridional Overturning Circulation (AMOC). Here, a light gray coloring is used to denote the two boxes whose salinities do not change, as well as all the arrows indicating terms which do not appear in the equations describing the dynamics of the 3-box model. Adapted from \citet{Alkhayuon2019}.}
\label{fig:Boxmodelsketch} 
\end{figure}%

\begin{figure}
\includegraphics[width = 0.8\linewidth]{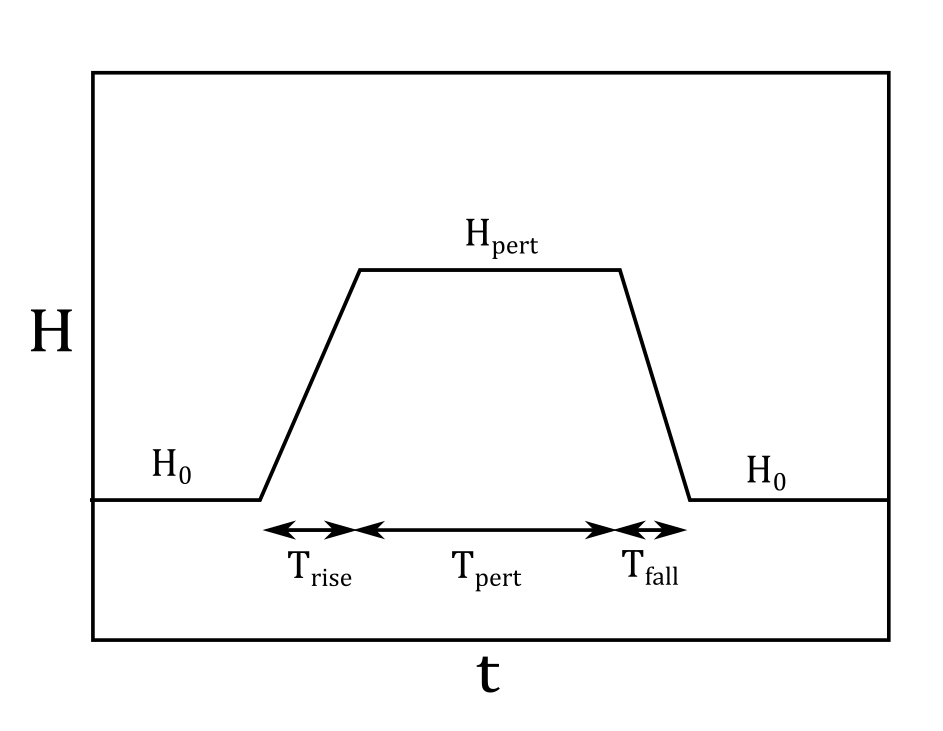}
\caption{ Schematic illustration of the piece-wise linear hosing function used to simulate the influx of fresh water. Adapted from \citet{Alkhayuon2019}.}
\label{fig:HosingSketch}
\end{figure}

\begin{figure}
\includegraphics[width = 0.8\linewidth]{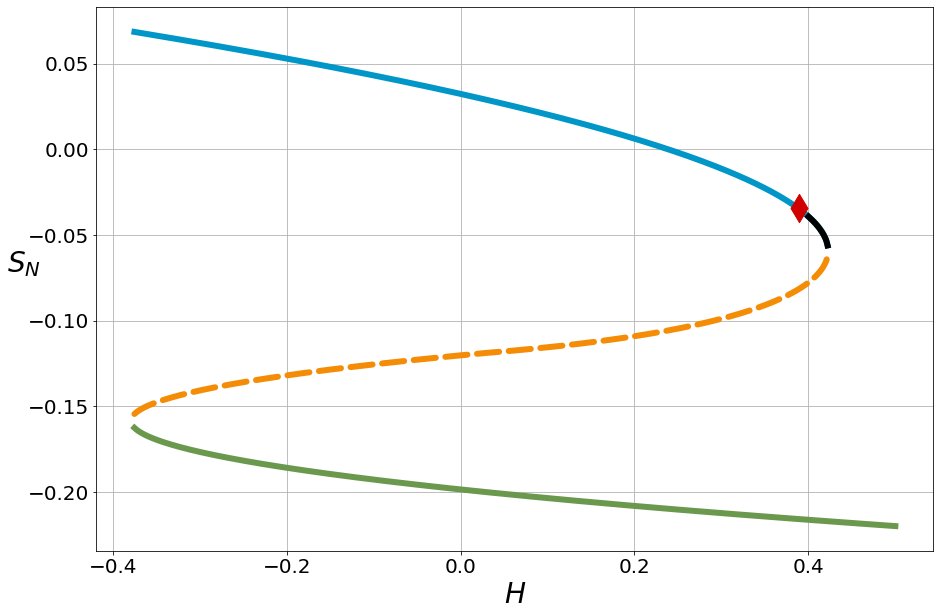}
\caption{ Bifurcation diagram for $S_N$, for the 3-box model of the AMOC. The dashed line denotes the unstable equilibrium branch. The red diamond denotes the location of the hopf-bifurcation.}
\label{fig:BifurcationDiagram}
\end{figure}

\section{Application to the global oceanic 3-box model} 
To determine the validity of the $\Upsilon$-indicator as a measure of stability, as well as its ability to detect different types of tipping points, we start by applying the method to the global oceanic 3-box model discussed by \citet{Alkhayuon2019}.
The 3-box model of \citet{Alkhayuon2019} is a simplification of the 5-box model of \citet{Wood2019} in which the salinity of the Southern Ocean (S) and the Bottom waters (B) is assumed to be approximately constant. The model thus consists of 5 separate boxes, of which only 3 boxes, namely the North Atlantic (N), Tropical Atlantic (T) and Indo-Pacific (IP) boxes have varying salinities $S$. A schematic illustration of the model is shown in Figure \ref{fig:Boxmodelsketch}.
See \citet{Alkhayuon2019}  or \citet{Wood2019} for a detailed exposition of the box model. We note that the parameters of the box model are tuned using the full complexity FAMOUS AOGCM model, with varying levels of CO$_2$. The parameters used in this paper are for the case $2\times$CO$_2$ as compared to pre-industrial times. \\
We denote salinity by $S_i$, the volume by $V_i$ and the fluxes by $F_i$, where $i\in \{N,T,S,IP,B\}$ denotes the respective boxes. \\ Let $\Gamma$ denote the AMOC flow defined by 
\begin{equation}
\label{eq:q_AMOCflow}
\Gamma = \lambda \left[\alpha(T_S-T_0)+\frac{\beta}{100}(S_N-S_S)\right]
\end{equation}
The model approximates a buoyancy-driven flow, with a transport proportional to the density difference between the boxes, assuming a linearized equation of state. The evolution equations for the salinities $S_N$ and $S_T$ are
\begin{widetext}
\begin{eqnarray}
\label{dS_N/dt_1}
    \frac{V_N}{Y} \frac{dS_N}{dt} &=& \Gamma(S_T-S_N) + K_N(S_T-S_N)-100 F_NS_0 \\
\label{dS_T/dt_1}
    \frac{V_T}{Y} \frac{dS_T}{dt} &=& \Gamma\left[\gamma S_S + (1-\gamma) S_{IP} -S_T\right] + K_S(S_S-S_T)+K_N(S_N-S_T)-100 F_TS_0 
\end{eqnarray}
for $\Gamma\geq 0$, and 
\begin{eqnarray}
\label{dS_N/dt_2}
    \frac{V_N}{Y} \frac{dS_N}{dt} &=& |\Gamma|(S_B-S_N)+K_N(S_T-S_N) - 100 F_NS_0 \\
\label{dS_T/dt_2}
    \frac{V_T}{Y} \frac{dS_T}{dt} &=& |\Gamma|(S_N-S_T) + K_S(S_S-S_T)+K_N(S_N-S_T)- 100 F_TS_0
\end{eqnarray}
\end{widetext}
for $\Gamma < 0$, where $S_B$ and $S_S$ are regarded as fixed parameters and $Y =  3.15 \times 10^7$, which converts the time unit from seconds to years. $S_0$ is a reference salinity, and $K_i$ are coefficients associated with the gyre strengths. We note that all the salinity values are given as perturbations from a background state, see Appendix A of \citet{Alkhayuon2019}\ for details on the transformation. Since the total salinity is assumed to be conserved, the salinity of the Indo-Pacific (IP) box, $S_{IP}$, can be computed from $S_N$ and $S_T$.\\
The values of the assorted parameters can be found in Table 1 and Table 2. \\
The fluxes, $F_N$ and $F_T$, are linear functions of the hosing function $H(t)$ which simulates the influx of fresh water. In the case of $2\times$CO$_2$ the fluxes are (see \citet{Wood2019})
\begin{eqnarray}
    F_N &=& 0.486\times10^6  + H(t)\;0.1311\times10^6 \\
    F_T &=& -0.997\times10^6 + H(t)\;0.6961\times10^6
\end{eqnarray}
where all fluxes are given in units of Sverdrup (Sv).\\
The values for the case of $1\times$CO$_2$ can be found in Table 5 of \citet{Alkhayuon2019}. \\
Figure \ref{fig:BifurcationDiagram} shows the bifurcation diagram for $S_N$; for $S_T$ we refer to \citet{Alkhayuon2019} The bifurcation diagram for the flow strength $\Gamma$ is qualitatively similar, since all other parameters in Eq. \ref{eq:q_AMOCflow} are kept constant. The diagram clearly shows that this is a bi-stable system with two stable equilibrium branches connected by an unstable branch. The upper equilibrium branch looses stability, not at the saddle-node bifurcation, but rather due to a Hopf-bifurcation, indicated by a red diamond in the diagram. Thus, part of the upper equilibrium branch, denoted in black, is in fact unstable.  
\\\\
To simulate the influx of fresh water we apply a time dependent, piece-wise linear  hosing function, $H(t)$ (see Figure \ref{fig:HosingSketch}), to equations (\ref{dS_N/dt_1})-(\ref{dS_T/dt_2}). Here
\begin{equation}
\label{H_pwl}
    H(t) = \begin{cases}
    H_0 & t < 0 \;,\\
    H_0 + \alpha(t) & t \in[0,T_{rise}] \;,\\
    H_{pert} & t-T_{rise} \in [0,T_{pert}] \;,\\
    H_{pert} - \beta(t) & t-T_{rise}-T_{pert} \in [0, T_{fall}]\;,\\
    H_0 & t \geq T_{rise} +T_{pert}+T_{fall} \;,\\
    \end{cases}
\end{equation}
where $\alpha(t)$ and $\beta(t)$ are linear functions ensuring continuity of $H(t)$. If we define the rise and fall rates, as
\begin{equation}
    r_{rise} = \frac{|H_{pert}-H_0|}{T_{rise}} \quad\text{ and } \quad r_{fall} = \frac{|H_{pert}-H_0|}{T_{fall}}
\end{equation}
then 
\begin{equation}
    \alpha(t) = r_{rise}t \quad \text{ and } \quad \beta(t) =  r_{fall}(t - T_{rise} - T_{pert})
\end{equation}
As demonstrated by \citet{Alkhayuon2019}, whether  the system undergoes a transition from one stable state to the other, is dependent not only on the value of $H_{pert}$, but on the rise and fall rates, $r_{rise}$ and $r_{fall}$, as well as the perturbation time $T_{pert}$. In particular, they demonstrate that even when $H_{pert}$ is above the bifurcation value that destabilizes the upper equilibrium branch, the system may still return to this equilibrium, provided $T_{fall}$ is short enough; a process which they termed \textit{avoided B-tipping}. In addition, they showed that if $T_{pert}$ is too short, the system will not tip, but return to the initial equilibrium branch.\\
In what follows, we will apply the \textbf{$\Upsilon$} indicator as described in the previous section to time series data generated by the 3-box model. We will separately study time series undergoing rate-, noise- and bifurcation-induced tipping, while attempting to assess the indicator's ability to gauge the stability of the time series as it approaches the tipping point. Before proceeding, we should clarify one point regarding noise-induced tipping, and what is meant by an early warning indicator in this context. Noise-induced tipping is inherently unpredictable, and hence one might conclude that any attempt at predicting such transitions is doomed to fail based on a single time series.
In contrast, assuming the underlying model is known, one could use ensembles of realizations to estimate the likelihood of noise-induced transitions. Examples of these statistical approaches are discussed in \citet{thompson_climate_2011}.
Although one cannot expect to develop an \textit{early} warning indicator for these types of transitions, one should at the very least be able to tell, from time series data, once such a transition has occurred, i.e., when the unstable equilibrium branch has been crossed and the system is approaching a different equilibrium. The objective should then be to develop an indicator that is able to identify this induced instability as soon as possible after the transition.  \\
Finally, we note that, while it is possible to extend ARMA fitting to multivalued time series data, we have chosen to not go down that route, and instead only apply the indicator to a single time series for the salinity values from the North Atlantic basin, $S_N$. The reason for choosing $S_N$ over $S_T$ is that within the 3-box model, the equilibrium branches of $S_N$ are that much further apart, making the transitions easier to see. Such a simplification might at first glance seem rather contrived, however we argue that, as the goal of any indicator is to be used on real-world time series data in which the connection to other time series is largely unknown, it is reasonable to only concentrate on one time series, despite the underlying system being multidimensional. 

\subsection{Bifurcation-induced Tipping}
\begin{figure}
    \centering
    \includegraphics[width =  \linewidth]{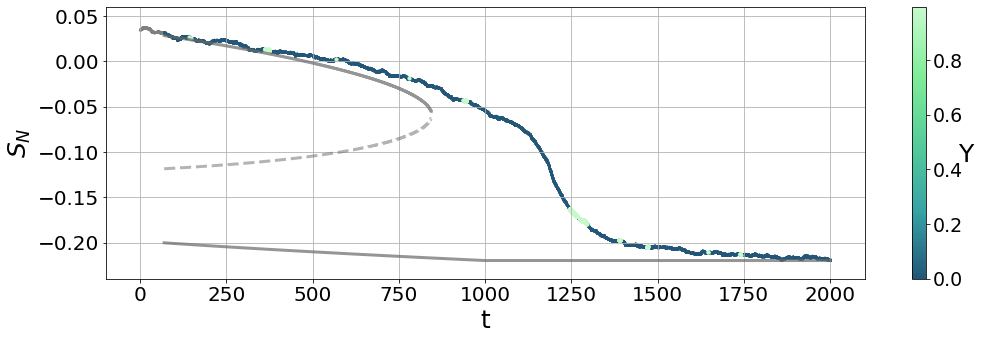}
    \caption{Bifurcation-induced tipping, color coded according to the value of \textbf{$\Upsilon$}  with window length, $\tau = 350$. The gray lines denote the equilibrium branches, with the dashed line corresponding to the unstable branch. We clearly see several brightly colored points corresponding to a high values of \textbf{$\Upsilon$}, which should be indicative of a high degree of instability and an approaching tipping point.}
    \label{fig:Btipping_Phaseportrait_colorplot}
\end{figure}

\begin{figure}
    \includegraphics[width = \linewidth]{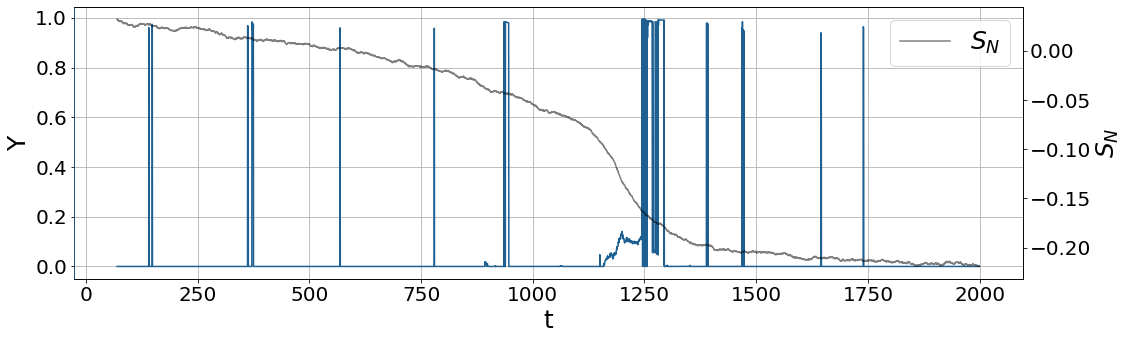}
    \caption{\textbf{$\Upsilon$} as a function of time for a time series of $S_N$ undergoing B-tipping.}
    \label{Btipping_Yplot}
\end{figure}

To induce B-tipping in the 3-box model, we gradually change $H(t)$ according to equation (\ref{H_pwl}), with $H_0=0$, $H_{pert} = 0.5$, $T_{rise}=1000$. This corresponds to an increase in the freshwater fluxes $F_T$ and $F_N$, corresponding to the flux into the tropical and North Atlantic boxes, by approximately $34\%$ and $13\%$, respectively. This, in turn, corresponds to roughly a 0.1-0.2 Sv increase, in line with freshwater "hosing" experiments of the North Atlantic \cite[e.g.][]{roche2014}. We let $T_{pert}$ go to infinity, such that $H(t)$ never returns to its initial value. As $H(t)$ changes, $S_N$ follows the upper equilibrium branch as sketched in Figure \ref{fig:BifurcationDiagram}, until it reaches the hopf-bifurcation (around $H = 0.4$), at which point the upper equilibrium branch becomes unstable, and $S_N$ starts approaching the lower equilibrium branch. We choose a window length of $350$ points corresponding to about 70 years. \\
Figure \ref{fig:Btipping_Phaseportrait_colorplot} shows the time series of $S_N$ color coded according to the value of \textbf{$\Upsilon$}, with brighter colors corresponding to higher values of \textbf{$\Upsilon$} and hence a greater degree of instability. Figure \ref{Btipping_Yplot} shows \textbf{$\Upsilon$} as a function of time, with clear peaks corresponding to brightly colored points in Figure \ref{fig:Btipping_Phaseportrait_colorplot}.\\ 

\begin{figure}
\centering
\subfloat(a){%
  \includegraphics[width=\linewidth]{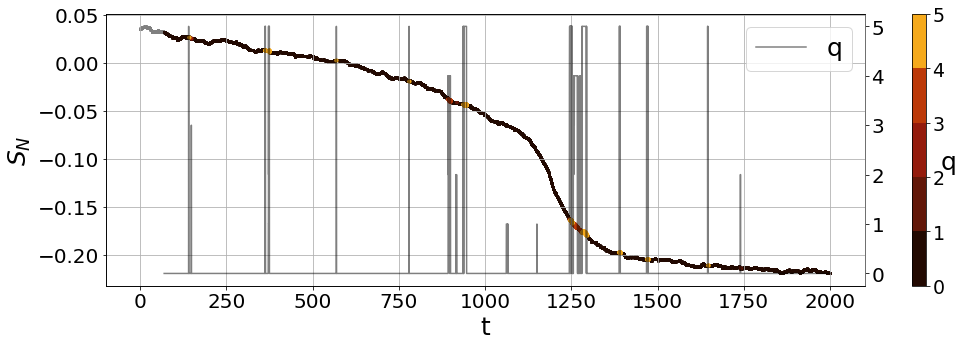}%
}
\subfloat(b){%
  \includegraphics[width=\linewidth]{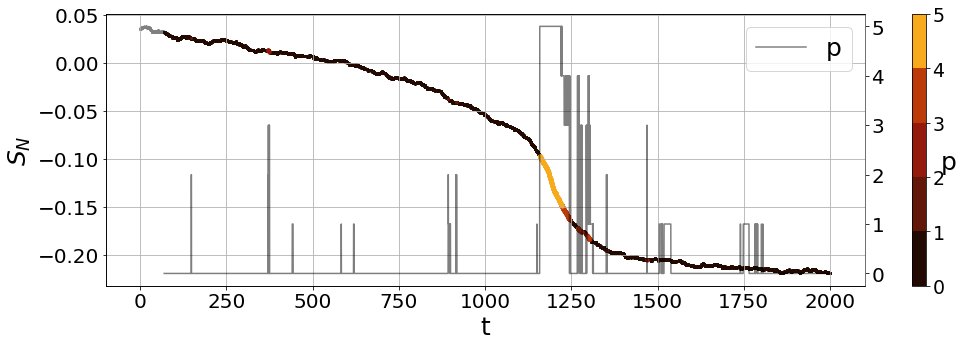}%
}
\caption{Bifurcation induced tipping of $S_N(t)$, color coded according to the value of the best-fit ARMA model orders (a) $q$ and (b) $p$ (scatter plot). The line plots additionally show the same values for $q$ and $p$ as functions of time in (a) and (b), respectively. }
\label{fig:Btipping_qpPlot}
\end{figure}

\begin{figure}
    \includegraphics[width = \linewidth]{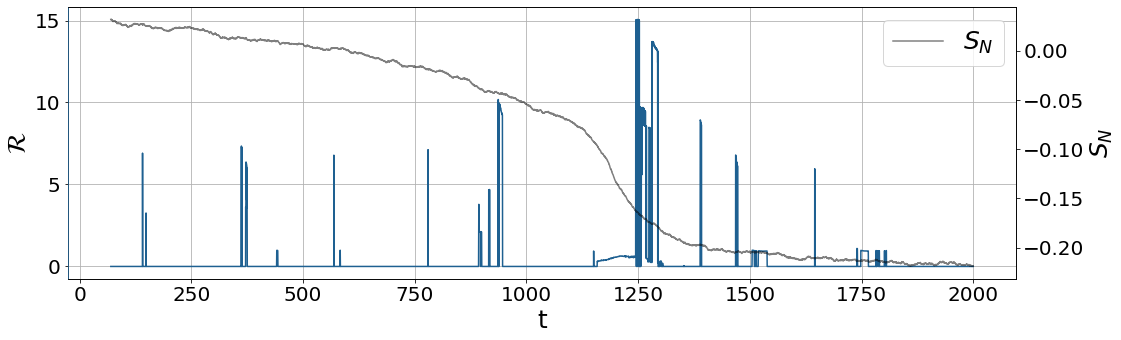}
    \caption{Plot of the persistence $\mathcal{R}$ (Eq. \ref{eq:persistence}) as a function of time for a time series of $S_N$ undergoing B-tipping.}
    \label{Btipping_Persistenceplot}
\end{figure}

It should be noted that low amplitude white noise is also applied to facilitate ARIMA model fitting. The noise intensity is kept small enough to avoid noise-induced tipping. \\
Figures \ref{fig:Btipping_Phaseportrait_colorplot} and \ref{Btipping_Yplot} clearly indicate that there are several points on the time series as it approaches the transition, which are deemed to have a high degree of instability. We further note that, although the result is not shown here, the high $\Upsilon$ values in Figures \ref{fig:Btipping_Phaseportrait_colorplot} and \ref{Btipping_Yplot} correspond to intervals for which 
$\Delta \text{BIC}_1(p,q)$ is negative, indicating that, as discussed previously, the ARMA(1,0) model would, when only considering BIC values, be the better fit, but it violates the auxiliary conditions, indicating a loss of stationarity. Hence, at these points ARMA(1,0) is excluded as a possible model, implying that ARMA(0,0) is the chosen base model.  
\\
In addition, we look at the order of the best-fit ARMA model, namely the $q$ and $p$ values, as well as the persistence, to gain further insight into the stability properties of the time series. Figure \ref{fig:Btipping_qpPlot} shows the time series of $S_N$ color coded according to the values of $q$ and $p$. When comparing with Figure \ref{fig:Btipping_Phaseportrait_colorplot}, this seems to indicate that the high values of \textbf{$\Upsilon$} appearing before the transition are primarily associated with an increase in the $q$-values. This is not unexpected, as it is primarily the change in the properties of the noise which is expected to give an indication of an approaching transition. Figure \ref{Btipping_Persistenceplot} shows the persistence plotted as a function of time $t$. We see a clear increase in the persistence directly preceding the tipping point around $t=1000$. \\
We make a final comment regarding Figure \ref{fig:Btipping_qpPlot} and its relation to our choice of ARMA(1,0) and ARMA(0,0) as base models. In \citet{Faranda2015} this choice was guided by the fact that for the time series under consideration the order, i.e. $p+q$, of the intervals was clustered around 1, and as the authors explicitly excluded pure moving-average processes, they concluded that {ARMA(1,0)} was the appropriate base model. However, from Figure \ref{fig:Btipping_qpPlot} we see that for the time series currently under consideration, the order is clustered around 0. This observation further strengthens the case for using ARMA(0,0) as an additional base model. We hypothesize that the dominance of ARMA(0,0) is related to the low degree of noise in the system, which makes the restoring force that returns the system to equilibrium less prominent, hence obscuring tendency of the random-walk to be attracted to a metastable state.

\begin{figure}
 \centering
\subfloat(a){%
  \includegraphics[width=\linewidth]{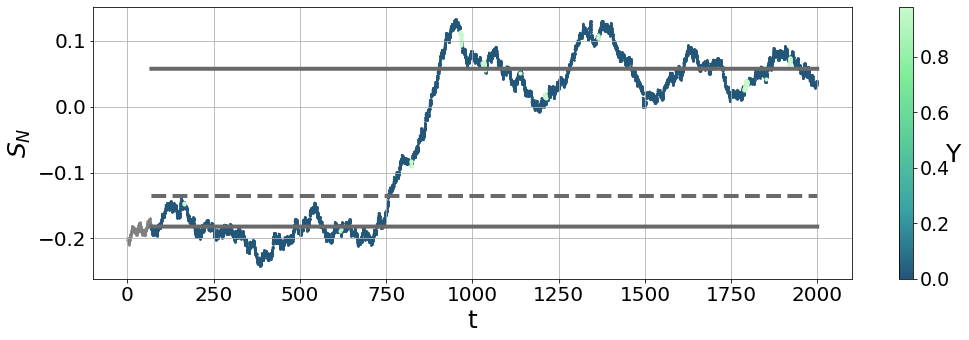}%
}
\subfloat(b){%
  \includegraphics[width=\linewidth]{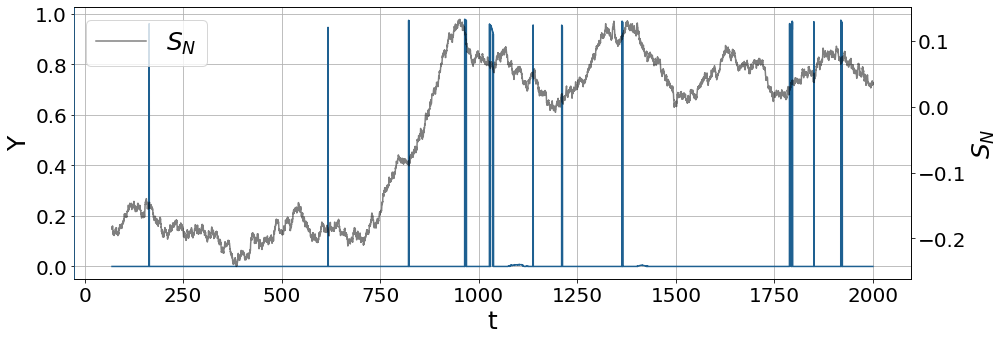}%
}
\caption{(a) Noise-induced tipping, color coded according to the value of \textbf{$\Upsilon$}. The gray lines denote the equilibria, with the dashed line denoting the unstable equilibrium branch. Transition from the lower to the upper equilibrium branch for $H=-0.25$, $\tau = 350$. (b) Plot of $\Upsilon$ as a function of time. Note how the peaks correspond to the brightly colored points in (a).}
\label{fig:Y_colorplot_Ntipping_Run5}
\end{figure}

\begin{figure}
 \centering
\subfloat(a){%
  \includegraphics[width=\linewidth]{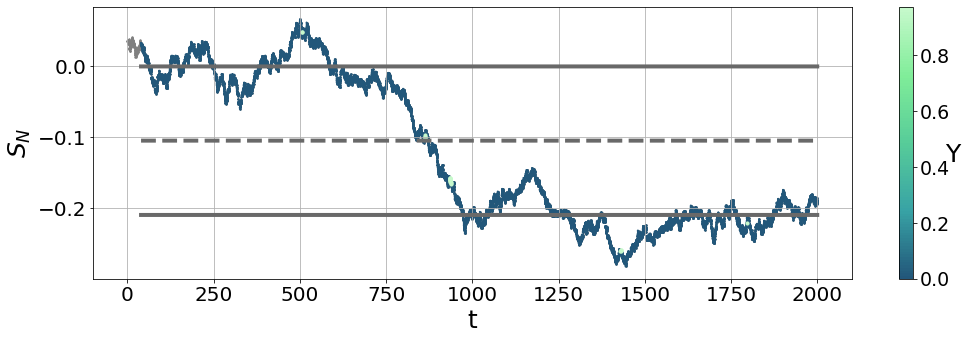}%
}
\subfloat(b){%
  \includegraphics[width=\linewidth]{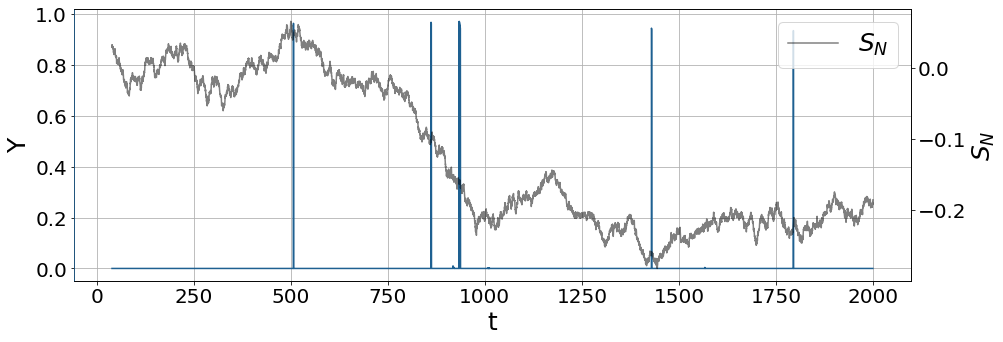}%
}
\caption{ (a) Noise-induced tipping, color coded according to the value of \textbf{$\Upsilon$}. The gray lines denote the equilibria, with the dashed line denoting the unstable equilibrium branch. Transition from the upper to the lower equilibrium branch for $H=0.24$, $\tau = 200$. (b) Plot of $\Upsilon$ as a function of time. Note how the peaks correspond to the brightly colored points in (a).}    
 \label{fig:Y_colorplot_Ntipping_Run6}
\end{figure}

\subsection{Noise-induced Tipping}

\begin{figure}
\centering
\subfloat(a){%
  \includegraphics[width=\linewidth]{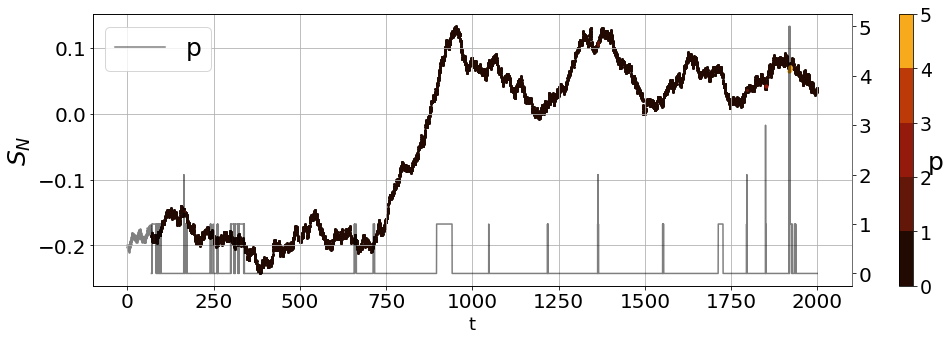}%
}
\subfloat(b){%
  \includegraphics[width=\linewidth]{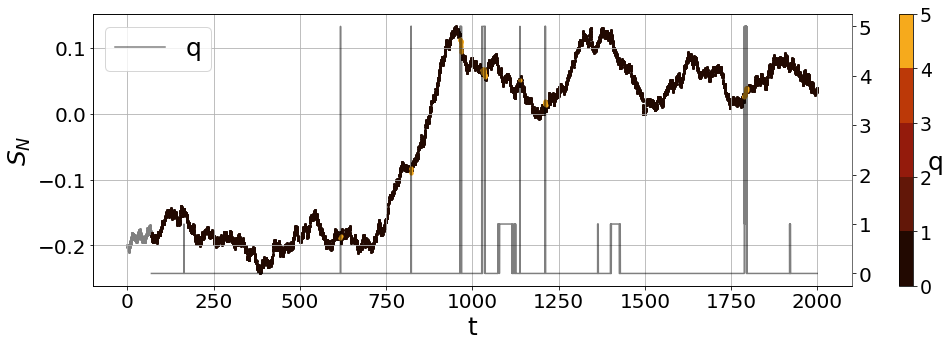}%
}
\caption{Noise-induced tipping of $S_N(t)$ for $H=-0.25$, $\tau = 350$, color coded according to the value of (a) $p$ and (b) $q$. For clarity we have also plotted is $p$ and $q$ as functions of time in (a) and  (b), respectively.}
\label{fig:Ntipping_qpPlot_Run5}
\end{figure}

\begin{figure}
\centering
\subfloat(a){%
  \includegraphics[width=\linewidth]{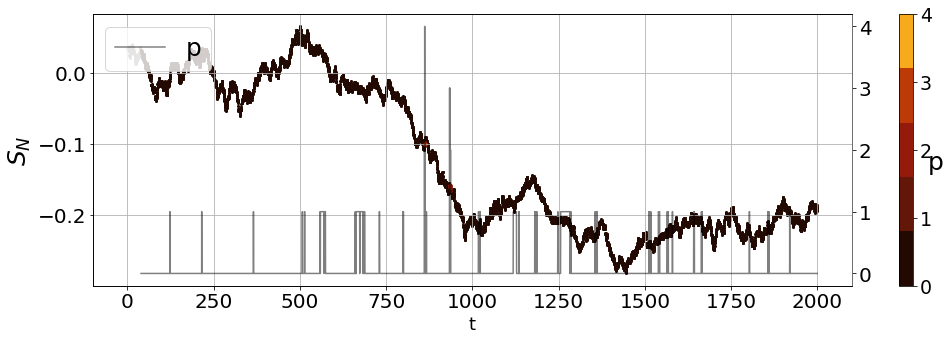}%
}
\subfloat(b){%
  \includegraphics[width=\linewidth]{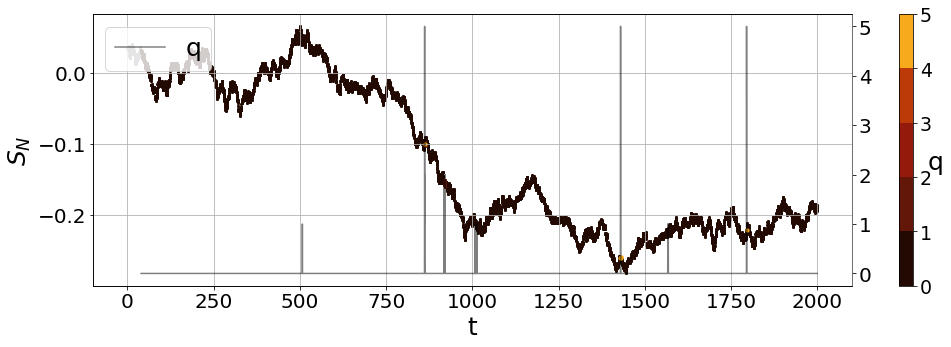}%
}
\caption{Noise-induced tipping of $S_N(t)$ for $H=0.24$, $\tau = 200$, color coded according to the value of (a) $p$ and (b) $q$. For clarity we have also plotted is $p$ and $q$ as functions of time in (a) and  (b), respectively.}
\label{fig:Ntipping_qpPlot_Run6}
\end{figure}

To induce N-tipping, we fix the hosing parameter $H$ and apply additive white noise to all the equations equally. The noise term is added equally to (\ref{dS_N/dt_1})-(\ref{dS_T/dt_2}), with the same noise amplitude in all cases. We look at transitions from the upper branch to the lower branch and \textit{vice versa}. In either case, it is convenient to choose a value for $H$ that is close to the bifurcation point, as the probability of transitioning is much higher in these regions, and hence one does not need high amplitude noise to induce transitions between the branches. \\
Figures \ref{fig:Y_colorplot_Ntipping_Run5} and \ref{fig:Y_colorplot_Ntipping_Run6} show two time series undergoing noise induced tipping, one going from the lower to the upper branch, while the other going the other way around. In the first case $H=-0.25$, while in the second $H=0.24$. The amplitude of the additive white noise is the same in both cases. For the window length $\tau$, we have chosen a length of 350 and 200 points, corresponding to about 70 and 41 years, respectively. The window length is chosen so that it is at most half as long as the transition time, which is taken to be the time for the system to arrive at the other equilibrium once it has crossed the unstable branch. Of course, when dealing with simulation data such as this, we have the advantage of knowing where the stable and unstable branches are, which is an advantage that anyone dealing with real-world data does not have. In principle one could use the clustering methods proposed by \citet{Kaiser2020} to approximate the window length, although this method also requires that one knows how many clusters, i.e., equilibrium states, one should look for. The clustering method works particularly well for noise induced transitions, as one can repeatedly induce transitions back and forth, to gain an ensemble of transitions, yielding a higher degree of accuracy. \\
In previous works, the choice of $\tau$ has largely been guided by a desire to ensure the stationarity of the time series intervals. However, as we are not requiring the individual time series segments to be stationary \textit{a priori}, we are permitted to use much longer time series intervals. In the world of ARIMA fitting a time series of length above 200 points would generally be considered a very long series, however, we should keep in mind that the sampling frequency of our simulated data is quite high; in fact, there are 5 points per time unit (i.e., year), yielding a total of 10000 points for the 2000 years of simulations. An interval consisting of 200 points corresponds to around 40 years, which is not an unreasonably long time interval for the dynamics of the AMOC. When fitting an ARIMA model to a time series, one wishes to avoid too long time series to avoid including events from the past that no longer have any relevance for the future. This, and not the inherent inaccuracy of the fit itself, is the primary reason for limiting the length of a time series. \\\\
Returning to Figures \ref{fig:Y_colorplot_Ntipping_Run5} and \ref{fig:Y_colorplot_Ntipping_Run6}, we note that there are a few brightly colored points indicating a high degree of instability. There are for example, in both cases, several points in the middle of the gap between the two stable branches, indicated by solid gray lines in the figure. This is consistent with the results of \citet{Kaiser2020}. In addition, for the transition from the lower to the upper branch, Figure \ref{fig:Y_colorplot_Ntipping_Run5}, there are several brightly colored points just after the system has reached the upper equilibrium branch. Although it is not so clear in the figure due to the presence of noise, any time $S_N$ returns to the upper equilibrium branch it initially overshoots and then oscillates around the equilibrium value with continuously decreasing amplitude (see Figure \ref{fig:RTipping_Y_notip_plot} for a clearer example of this behavior). This is probably due to the presence of an unstable limit cycle, and the aforementioned sub-critical hopf bifurcation. Hence, we see it as an encouraging sign that the indicator seems to be able to identify these points as well. We further note that, although the result is not shown, the high $\Upsilon$ value points in figure \ref{fig:Y_colorplot_Ntipping_Run5} and \ref{fig:Y_colorplot_Ntipping_Run6} correspond to points where $\Delta_1 \text{BIC}(p,q)$ is negative, as was the case for the B-tipping example in the previous section.\\ 
Looking at the $p$ and $q$ values in Figures \ref{fig:Ntipping_qpPlot_Run5} and \ref{fig:Ntipping_qpPlot_Run6}, it is clear that high values of \textbf{$\Upsilon$} correspond to high values of $q$, while the connection between $p$ and \textbf{$\Upsilon$} remains uncertain. However, we note that the high $\Upsilon$ values appearing around the transition correspond to high values of both $p$ and $q$, and consequently also of persistence (result not shown).

\begin{figure}
    \centering
    \includegraphics[width = \linewidth]{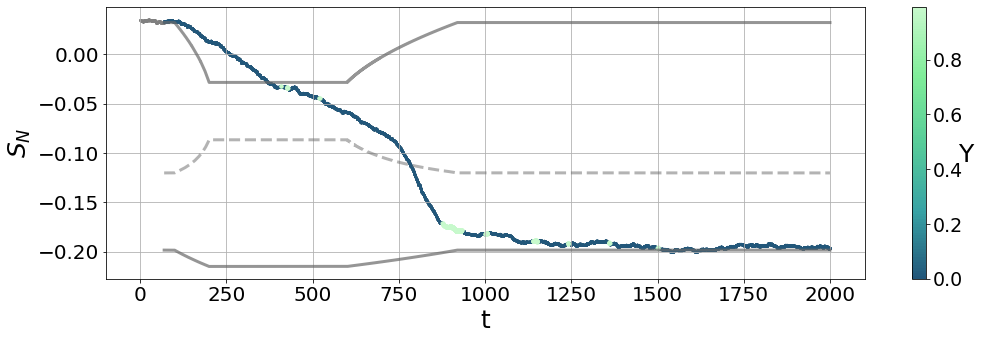}
    \caption{Rate-induced tipping of $S_N$, color coded according to the value of \textbf{$\Upsilon$}. The moving equilibria are plotted in gray, with the dashed line denoting the unstable branch. Compare this figure to Figure \ref{fig:Rtipping_q_plot}, which shows the same time series, but color coded according to the value of $q$.}
    \label{fig:RTipping_Yplot}
\end{figure}

\begin{figure}
    \centering
    \includegraphics[width = \linewidth]{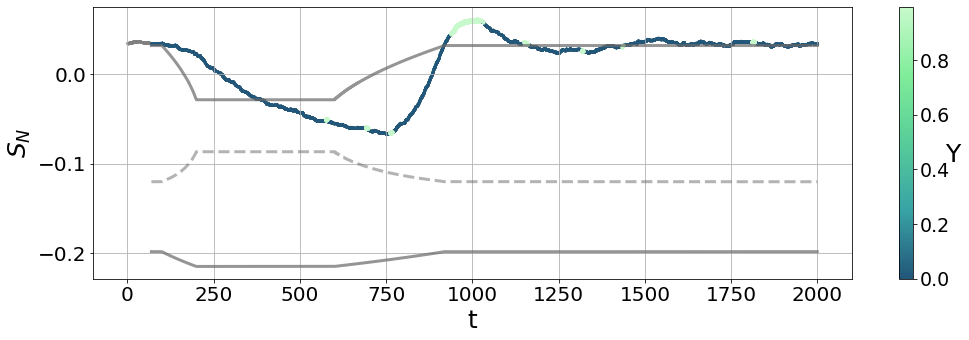}
    \caption{$S_N$ as a function of time, color coded according to the value of \textbf{$\Upsilon$} for $T_{fall}=280$. With these parameter values, the system does not tip, but returns to the upper equilibrium branch after some time. Note that the system initially overshoots the stable branch upon return. This is probably due to the presence of the unstable limit cycle. The equilibrium branches are plotted in gray, with the dashed line denoting the unstable branch.}
    \label{fig:RTipping_Y_notip_plot}
\end{figure}

\begin{figure}
\centering
\subfloat(a){%
  \includegraphics[width=\linewidth]{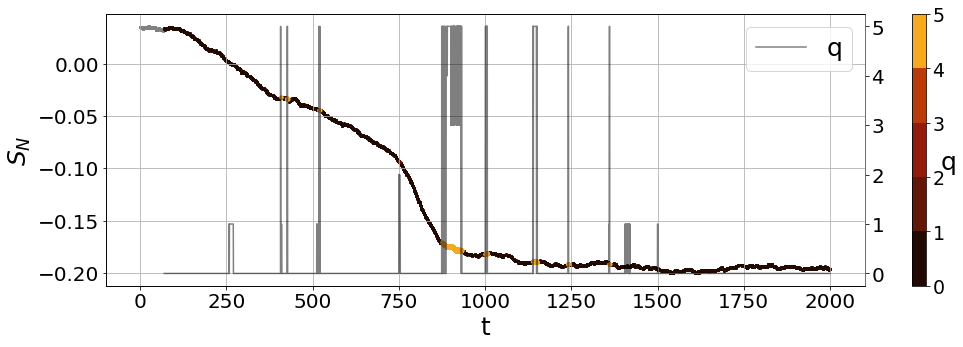}%
}
\subfloat(b){%
  \includegraphics[width=\linewidth]{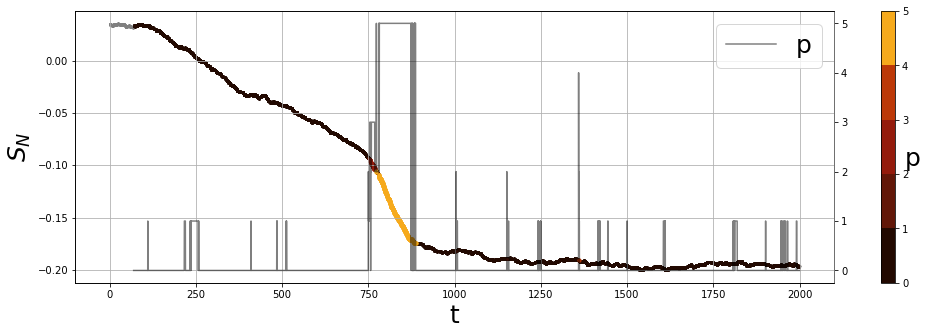}%
}
\caption{Rate-induced tipping of $S_N(t)$, color coded according to the value of (a) $q$  and (b) $p$. The value for $q$ and $p$ are also plotted as functions of time in (a) and (b), respectively. }
\label{fig:Rtipping_qp_plot}
\end{figure}

\begin{figure}
\centering
\subfloat(a){%
  \includegraphics[width=\linewidth]{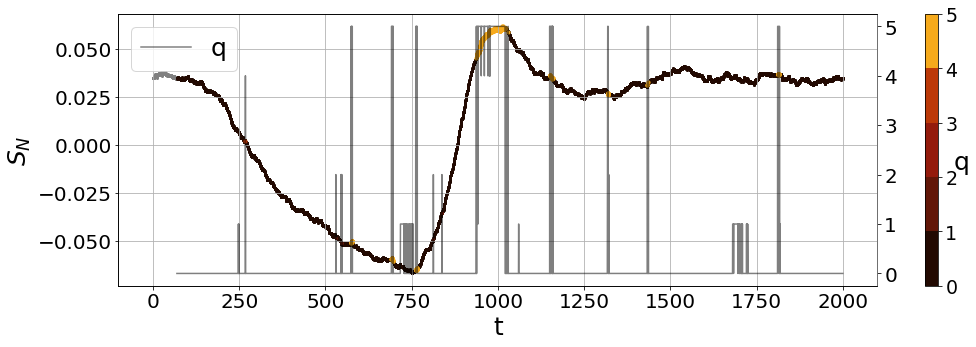}%
  }
\subfloat(b){%
  \includegraphics[width=\linewidth]{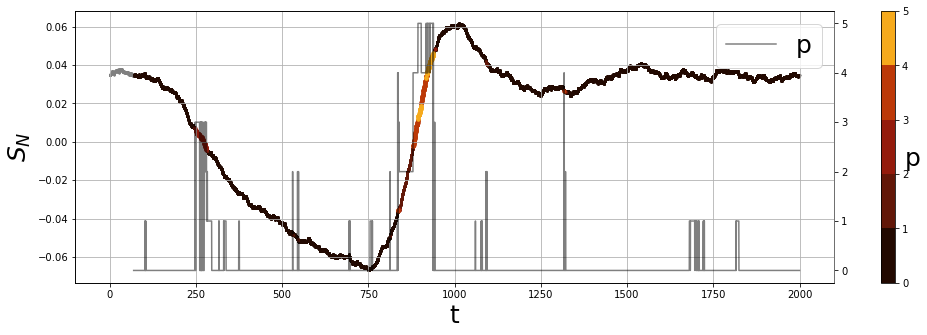}%
  }
\caption{$S_N$ as a function of time, color coded according to the value of (a) $q$  and (b) $p$ , for $T_{fall} = 280$. For these parameter values, the system does not tip, but returns to the initial equilibrium after some time $t$. For clarity, $p$ and $q$ are also plotted as functions of time in (a) and (b), respectively. It is instructive to compare these plots to Figure \ref{fig:RTipping_Y_notip_plot}.}
\label{fig:Rtipping_qp_plot_notip}
\end{figure}

\begin{figure}
    \centering
    \includegraphics[width =\linewidth]{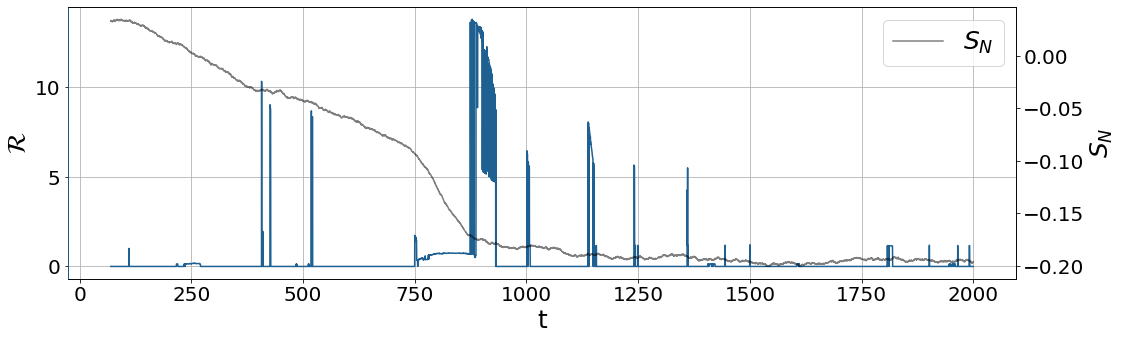}
    \caption{Persistence of a time series undergoing rate-induced tipping, plotted as a function of time. The underlying series is the time series shown in Figure \ref{fig:RTipping_Yplot}. We see several high persistence values, corresponding with a high value for the order, $q+p$ (compare with Figure \ref{fig:Rtipping_qp_plot}), appearing before the potential tipping point around $t=500$.}
    \label{fig:RTipping_Tipping_persistence_plot}
\end{figure}

\subsection{Rate-induced Tipping}

To induce R-tipping we fix $H_{pert}$ below the bifurcation value, ensuring that both equilibria still exist and are stable, and vary $T_{fall}$. We set $T_{rise} = 100$ and $T_{pert}=400$, while $H_{pert} = 0.37$. This corresponds to an increase in the freshwater fluxes $F_T$ and $F_N$, corresponding to the flux into the tropical and North Atlantic boxes, by approximately $25\%$ and $10\%$, respectively. Next, we observe that for $T_{fall}= 280$ the system returns to the upper equilibrium branch, while for $T_{fall} = 320$, the system transitions to the lower branch. The transition happens even though the bifurcation boundary has not been crossed. Again, we note that some additive white noise has been applied to allow for ARIMA fitting. \\
Figure \ref{fig:RTipping_Yplot} shows a time series undergoing rate-induced tipping, with the color coding corresponding to the values of \textbf{$\Upsilon$}. Again, we have chosen $\tau = 350$ points, corresponding to 70 years. We see several brightly colored points, indicating a high degree of instability, before the system transitions. These points occur initially as the system approaches the unstable branch (between approximately $t=350$ and $t=500$). These points do not appear for the time series that does not tip,  Figure \ref{fig:RTipping_Y_notip_plot}, despite the fact that within this time interval, the two time series are virtually identical, and could therefore be an indication of an approaching tipping point. However, again looking at Figure \ref{fig:RTipping_Y_notip_plot} we see some brightly colored points, corresponding to large $\Upsilon$, in the interval $t=600$ to $t=750$, and it is unclear what approaching instability these points would be indicative of, and thus might be regarded as false signals.\\
Looking at Figure \ref{fig:Rtipping_qp_plot}, it becomes clear that the high values of \textbf{$\Upsilon$} found in Figure \ref{fig:RTipping_Yplot} correspond to high values of $q$, while a comparison with Figure \ref{fig:RTipping_Tipping_persistence_plot}, gives the same indication for the persistence. In other words, high values of \textbf{$\Upsilon$} primarily correspond to high values of persistence and $q$.\\
From Figure \ref{fig:RTipping_Y_notip_plot}, we can also see how the indicator correctly identifies the unstable limit cycle, which we have argued causes the overshoot when returning to the upper equilibrium branch. Figure \ref{fig:Rtipping_qp_plot_notip} shows the same time series as in Figure \ref{fig:RTipping_Y_notip_plot}, color coded according to the values of $q$ and $p$. While high values of $q$ seem to be associated with increased instability, the high values of $p$ primarily occur as the system returns to the equilibrium. We would therefore suggest that high values of the autoregressive order, $p$, should be interpreted as an indication that the system is following a moving equilibrium branch. 
Comparing Figures \ref{fig:RTipping_Tipping_persistence_plot} and \ref{fig:Rtipping_q_plot} it becomes clear that the points with high $q$ value around $t=1000$, correspond to particularly high values of persistence, even when compared to other points of similar order. We also note that, as in the previous two tipping scenarios, the high $\Upsilon$ values, or equivalently high $p$ values, \\
We end this section with a brief comment on the rate-induced tipping example presented in this section. In this example the system is, as it undergoes rate-induced tipping, approaching a bifurcation boundary. It would be instructive to study a case in which this is not the case to ensure that the detected instability is not merely due to the approaching bifurcation boundary. However, as one would need to look at different model examples than those presented here, this is outside the scope of the current work.

\section{Comparison with Other Early Warning Indicators}
As briefly alluded to in the introduction, it is well established that bifurcation-induced tipping is generally preceded by an increase in lag 1 autocorrelation and variance (\citet{Lenton2012}, \citet{Dakos2012}  , \citet{boers_observation-based_2021}). 
The intuition behind this is that as the system approaches a bifurcation point, the potential well flattens out, reducing the speed at which the system recovers from a perturbation, so called "critical slowing down", which should manifest as an increase in the variance and autocorrelation of the time series. However, the variance and autocorrelation might also increase for other reasons, in particular if the properties of the noise changes. What happens to the autocorrelation and variance when the system approaches a rate-induced tipping point is thus far unclear, although it is conceivable that the "critical slowing down" hypothesis still holds for this type of tipping, see \citet{RitchieSieber2016}. Obviously, it does not hold true for time series undergoing purely noise induced tipping, as there is no change in the potential well. However, the autocorrelation and variance of the time series will dramatically change as the system crosses the unstable equilibrium branch and enters a different potential well. \\
In what follows, we will compare these classical indicators to the \textbf{$\Upsilon$} indicator for rate-induced and bifurcation-induced tipping in the AMOC 3-box model. It is instructive to just look at the part of the time series prior to the transition, as in general one wishes to be able to detect early signs of the transition \textit{before} it happens. For the time series undergoing bifurcation-induced tipping (Figure \ref{fig:Btipping_Phaseportrait_colorplot}) we chose a segment consisting of the points between approximately $t=200$ and $t=1100$. For the time series undergoing rate-induced tipping (Figure 13), we choose a  segment consisting of the points between $t=200$ and $t=700$. This segment is in all probability too long, meaning that it also contains the transition itself, as opposed to only points prior to the transition. However, this is the inherent difficulty with rate induced tipping; there is currently no way to analytically determine \textit{when} the transition happens, and one largely has to guess. Based on Figures  \ref{fig:RTipping_Yplot} and \ref{fig:RTipping_Y_notip_plot}, one could potentially conclude that the tipping point is found somewhere between $t=400$ and $t=600$, but this is pure guess work. For this reason we have included points up until $t=700$.\\\\
 Given a set of measurements $Y_1,Y_2,\cdots, Y_N$ the sample variance is defined as 
\begin{equation}
    \sigma^2 = \frac{1}{N}\sum_{i=1}^{N}\left(Y_i - \overline{Y}\right)^2
\end{equation}
while the lag k autocorrelation is given by
\begin{equation}
   \label{eq:AC}
    \text{r}_k = \frac{1}{N \sigma^2}\sum_{i=1}^{N-k}\left(Y_i - \overline{Y}\right)\left(Y_{i+k} - \overline{Y}\right)
\end{equation}
where $\overline{Y}$ denotes the sample mean of the series $Y_1,Y_2,\cdots, Y_N$ (see for example chapter 2 of \citet{box_jenkins2008}). Although time does not enter explicitly in the formulas, it is assumed that the measurements are taken at regular intervals. \\\\
When computing the variance and autocorrelation it is essential that the signal is properly detrended; otherwise any trend will immediately obscure the relevant dynamics. As for the \textbf{$\Upsilon$} indicator, one generally employs a rolling window approach, with an appropriately chosen window length $\tau$. \citet{Lenton2012} demonstrated that detrending can be done within each time window, as opposed to on the whole time series at once, without significantly changing the result. We have chosen this same approach, using linear detrending, as opposed to quadratic or higher order detrending methods, to remove the trend. The window length $\tau$ was set to 350 points, corresponding to 70 years. \\
\begin{figure}
    \centering
    \includegraphics[width =\linewidth]{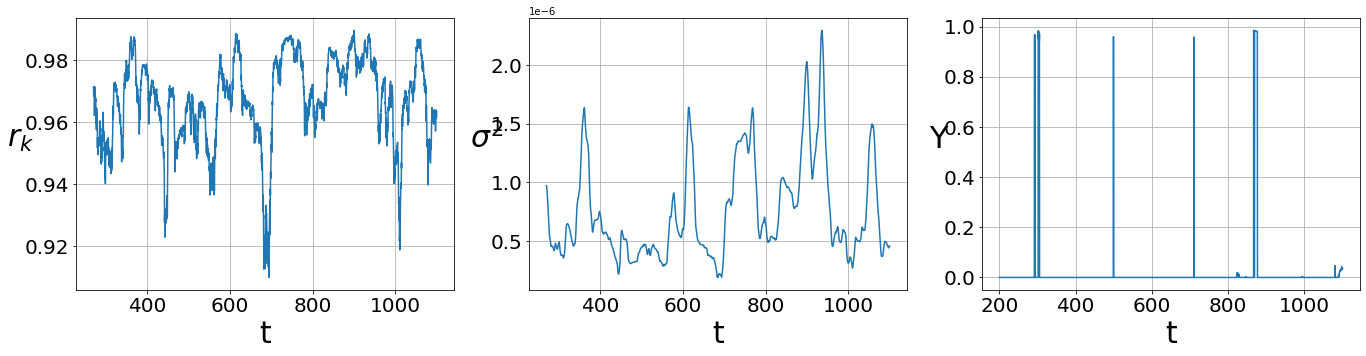}
    \caption{Autocorrelation, Variance and \textbf{$\Upsilon$} plotted as functions of time for a time series undergoing B-tipping. The increase in the variance as one approaches the tipping point is clear, while the increase in autocorrelation is less clear. }
    \label{fig:Comparison_ACVarY_Btipping}
\end{figure}

\begin{figure}
    \centering
    \includegraphics[width =\linewidth]{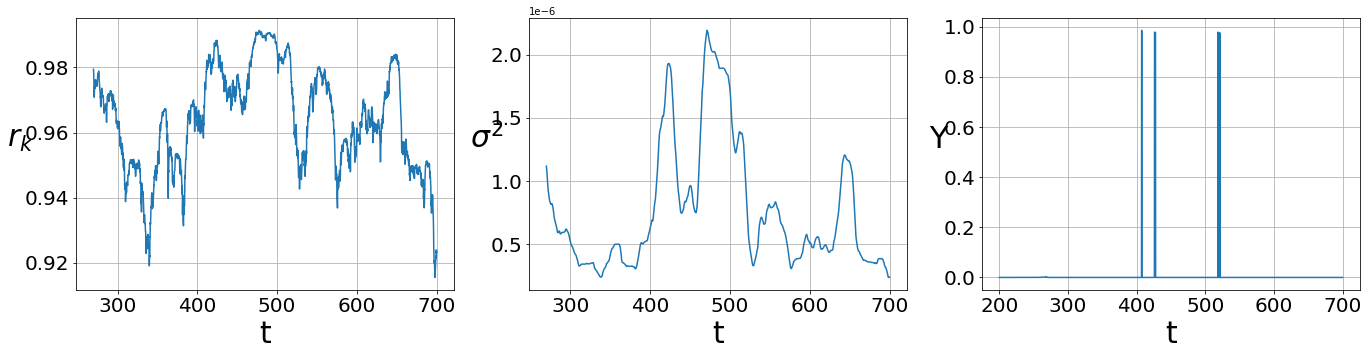}
    \caption{Autocorrelation, Variance and \textbf{$\Upsilon$} plotted as functions of time for a time series undergoing R-tipping. Assuming that the tipping point is around t=450, one can clearly see an increase in both autocorrelation and variance prior to the tipping point.}
    \label{fig:Comparison_ACVarY_Rtipping}
\end{figure}

\begin{figure}
    \centering
    \includegraphics[width =\linewidth]{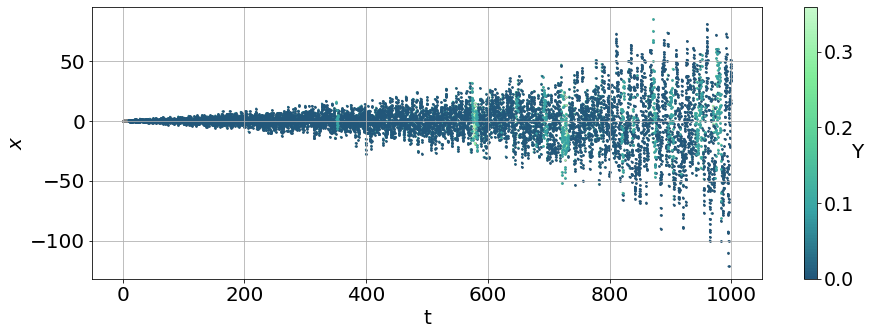}
    \caption{Time series with colored noise but no tipping points, color coded according to the value of \textbf{$\Upsilon$}.}
    \label{fig:Comparison_ColoredNoise_Y_plot}
\end{figure}
Figures \ref{fig:Comparison_ACVarY_Btipping} and \ref{fig:Comparison_ACVarY_Rtipping} show the autocorrelation, variance and \textbf{$\Upsilon$} plotted as functions of time. The peaks in \textbf{$\Upsilon$} preceding the transition are clear, as is the increase in variance and autocorrelation, at least in the case of R-tipping, provided the tipping point is approximately at $t=450$. For B-tipping, there appears to be a clear increase in the variance preceding the tipping point, provided the tipping point happens around $t=850$  (see Figure \ref{fig:Btipping_Phaseportrait_colorplot} for comparison). The expected increase in autocorrelation is, however, less clear.  \\ It is possible that the high degree of autocorrelation in the 3-box model, as observed in Figures \ref{fig:Comparison_ACVarY_Btipping} and \ref{fig:Comparison_ACVarY_Rtipping} is correlated to the frequent failure of the ARMA(1,0) model, whereby failure we mean that the autoregressive coefficent, sometimes referred to as the AR1 coefficient, violates the stationarity condition, and resulting in ARMA(1,0) being excluded as a possible candidate model. \\\
As already noted, the upper equilibrium branch does not lose stability due to a saddle node bifurcation, but rather loses stability due to a sub-critical Hopf bifurcation. It is possible that classical indicators are struggling to pick up on this. Furthermore, the noise amplitude is kept low to avoid noise-induced tipping, which might make it difficult for the indicators to pick up on changes in the dynamics.
\\

\begin{figure}
    \centering
    \includegraphics[width= \linewidth]{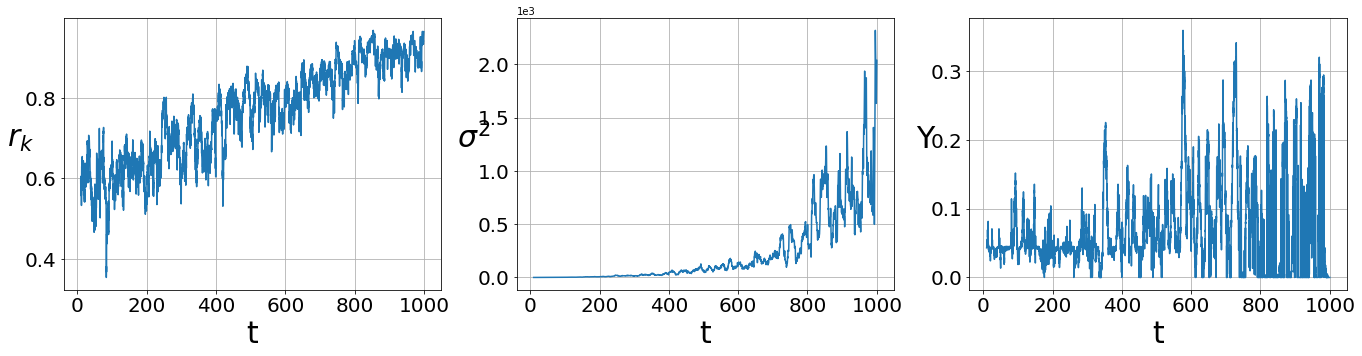}
    \caption{Autocorrelation, Variance and \textbf{$\Upsilon$} plotted as functions of time for a time series with colored noise but no tipping points. All three indicators show a dramatic increase, falsely suggesting an upcoming tipping point.}
    \label{fig:Comparison_ColoredNoise_ACVarY}
\end{figure}

\begin{figure}
\centering
\subfloat(a){%
  \includegraphics[width=\linewidth]{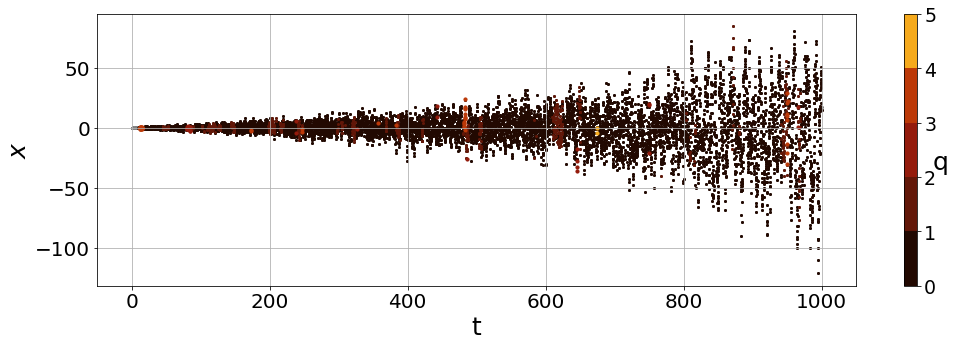}%
  }
\subfloat(b){%
  \includegraphics[width=\linewidth]{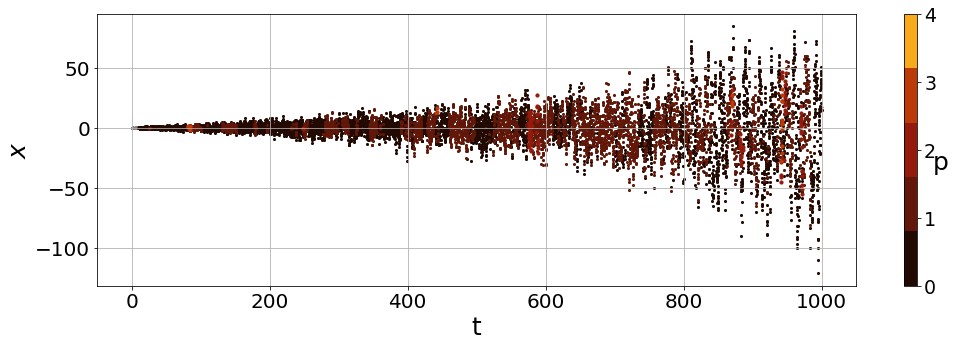}%
  }
\caption{Time series with colored noise and no tipping points, corresponding to equation (\ref{eq:ColoredNoise}), color coded according to the value of (a) $q$  and (b) $p$.}
\label{fig:ColoredNoise_qp_plot}
\end{figure}

\begin{figure}
    \centering
    \includegraphics[width = \linewidth]{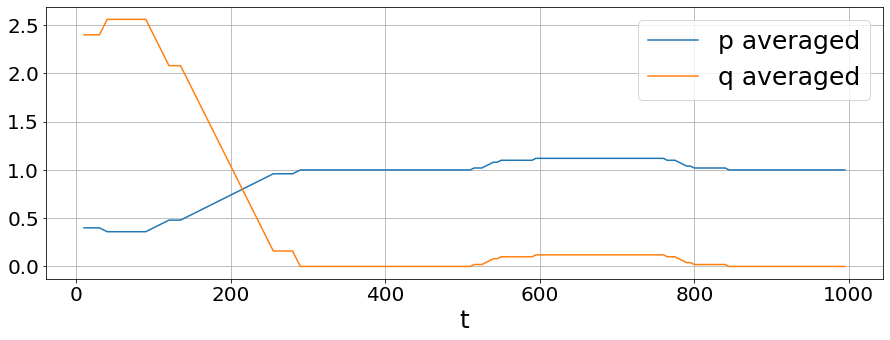}
    \caption{The values of $p$ and $q$ for the colored noise time series, averaged with a window length of 50 points, corresponding to 25 non-dimensional time units.}
    \label{fig:ColoredNoise_avergaed_pq}
\end{figure}
The autocorrelation and variance of a time series can increase for reasons that have nothing to do with an approaching tipping point. Hence, we wish to see how the \textbf{$\Upsilon$} indicator responds to colored noise, whose variance and autocorrelation increases with time $t$. To this end, we construct an artificial time series of the form
\begin{equation}
\label{eq:ColoredNoise}
\frac{dx}{dt} = -5x +\xi(t)
\end{equation}
where $\xi(t)$ is autocorrelated colored noise. $\xi(t)$ is in effect modelled as an ARMA(1,0) process whose AR1 coefficient increases linearly in time. In addition, the variance of this process also increases linearly in time. This is equivalent to the example presented in \citet{boers_observation-based_2021}. Applying the \textbf{$\Upsilon$} indicator to this time series yields the result shown in Figure \ref{fig:Comparison_ColoredNoise_Y_plot}. Figure \ref{fig:Comparison_ColoredNoise_ACVarY}  shows a comparison between the autocorrelation, variance and value of \textbf{$\Upsilon$} for the same time series. All three indicators show a dramatic increase, despite there being no approaching tipping point. However, looking at the plot of the time series when color coded according to the values of $p$ and $q$, Figure \ref{fig:ColoredNoise_qp_plot},  a curious pattern emerges: the increase in \textbf{$\Upsilon$} is largely associated with increased $p$ value. Looking at Figure \ref{fig:ColoredNoise_avergaed_pq} the trend becomes even clearer: here we have computed the rolling average of the $p$ and $q$ values with a window length of 50 points corresponding to 25 non-dimensional time units. We see that while the average value of $q$ goes towards zero for large $t$, the average value of $p$ settles around one. The general trend is independent of the choice of window length, provided the window length is between 30 and 300 points. 
\\ This behavior is unlike what was observed for the 3-box model. The high values of \textbf{$\Upsilon$} were associated with a high value of $q$. We thus argue that high values of $q$ were associated with increased instability, while high values of $p$ were more indicative of the system following a moving equilibrium. \\
Thus, one would, through the distinction between $q$ and $p$ values, potentially have a way of distinguishing the effect of colored noise from real early warning signals. However, it is conceivable that the result for the artificial colored noise time series is a consequence of how we have constructed the colored noise, so further studies on this are warranted. \\\\
Finally, we note that the constructed colored noise time series is a very artificial example of colored noise, as the noise amplitude increases by a probably unrealistic amount, and when applied to any reasonable time series it would obscure the dynamics altogether. This is to say that although we can likely assume that the noise in real-world data is autocorrelated, it will be much more subtle, and not result in equally high values of \textbf{$\Upsilon$}.

\section{Application to simulation data from {CESM2}}
So far, we have only applied the dynamic stability indicator to data from a very simplified 
model. The actual ocean has many more degrees of freedom and the response could be quite different. Nevertheless, it is of interest to see how the indicator responds when applied to such a system. To this end,
we employ 
data from the earth systems model CESM2, under two climate scenarios: one in which the atmospheric CO$_2$ concentration is abruptly doubled and another in which it is abruptly quadrupled. 
Both simulations were initialized using a pre-industrial control run ($piControl$) and then run for 500 years. The CO$_2$ was then increased, at $t=6000$ months. The data was saved at monthly intervals and the seasonal cycle was removed prior to the analysis. Such an abrupt change in CO$_2$ represents an extreme forcing, and contrasts with the ramped-up hosing employed with the idealized model. However, the oceanic response is not instantaneous, but requires 2-3 decades for freshwater to circulate in the model's sub-polar gyre \citep{Madan2022}. We consider this more hereafter.

\subsection{Abrupt $\mathbf{4\times CO_2}$}
\begin{figure}
    \centering
    \includegraphics[width = \linewidth]{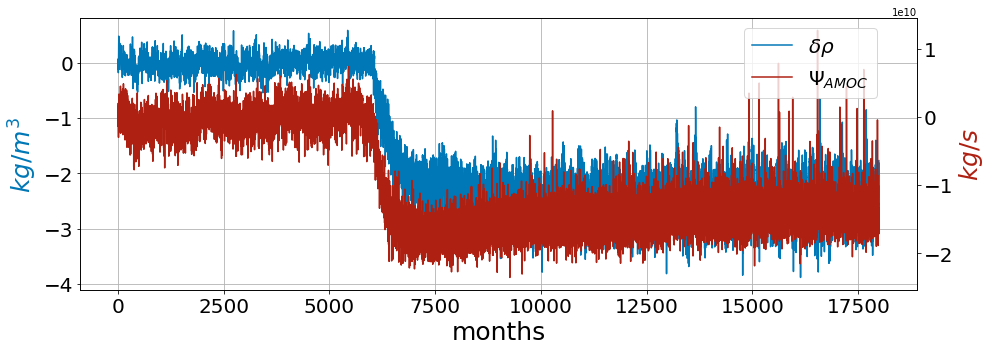}
    \caption{CESM2 model with abrupt $4\times \text{CO}_2$, where the monthly density difference (blue) is plotted together with the maximum AMOC flow strength (red). Note that the CO$_2$ was increased at t=6000 months.}
    \label{fig:NORESM_4XCO2_timeseries}
\end{figure}

\begin{figure}
    \centering
    \includegraphics[width = \linewidth]{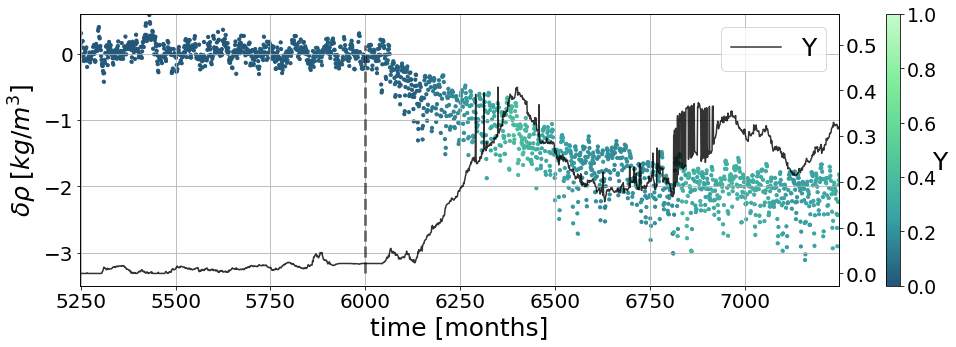}
    \caption{Time series of monthly density changes for abrupt $4\times \text{CO}_2$, color coded according to the value of \textbf{$\Upsilon$}. The window length is 250 points, corresponding to exactly 20 years. The dashed line indicates the point when the CO2
concentration abruptly changes.}
    \label{fig:NORESM_4XCO2_Y_colorplot}
\end{figure}

\begin{figure}
\centering
\subfloat(a){%
  \includegraphics[width=\linewidth]{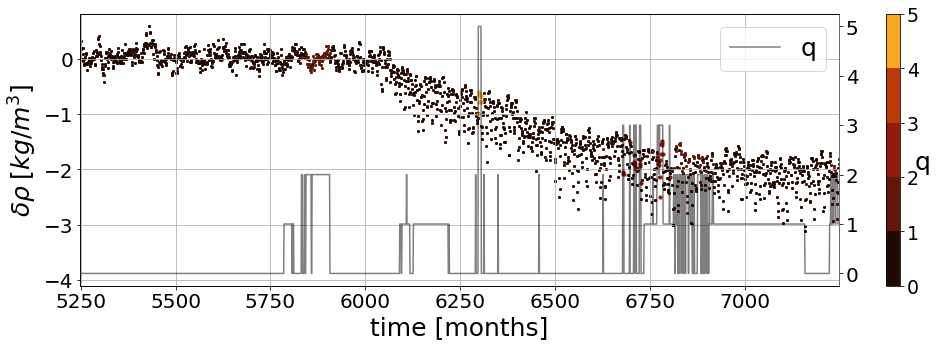}%
  }
\subfloat(b){%
  \includegraphics[width=\linewidth]{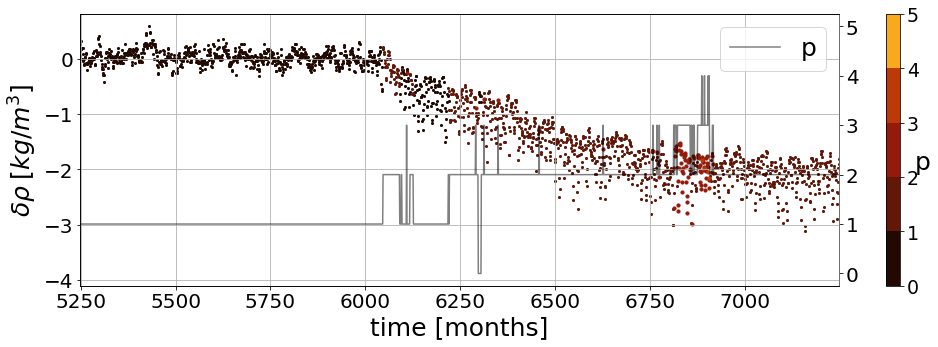}%
  }
\caption{Time series of monthly density changes for abrupt $4\times \text{CO}_2$, color coded according to the value of (a) $q$  and (b) $p$. The value for $q$ and $p$ are also plotted as functions of time in (a) and (b), respectively. }
\label{fig:NorESM_4XCO2_qp_plot}
\end{figure}

\begin{figure}
    \centering
    \includegraphics[width = \linewidth]{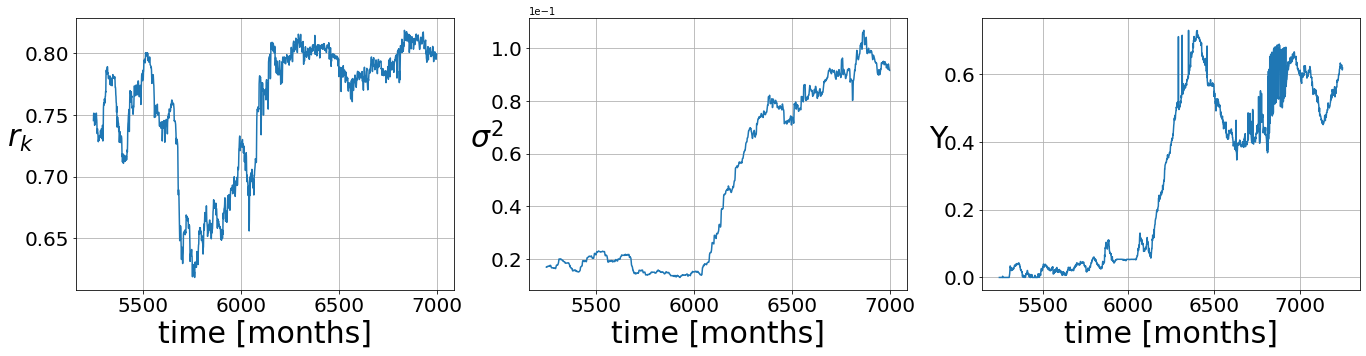}
    \caption{Autocorrelation, variance and \textbf{$\Upsilon$} plotted as functions of time for the case of abrupt $4\times \text{CO}_2$.}
    \label{fig:NorESM_4XCO2_AC_Var_Y_plot}
\end{figure}

The time series of a monthly-mean density difference, $\delta\rho$, and AMOC strength, $\psi_{AMOC}$, are shown in Figure \ref{fig:NORESM_4XCO2_timeseries}  for the case of abrupt $4\times \text{CO}_2$.
The density difference, a measure dynamically linked to the AMOC strength (\citet{Madan2022}), is calculated from the difference in surface densities averaged in boxes to the north and south of the North Atlantic Current. The surface density is calculated using the thermodynamic equation of state of seawater as per UNESCO 1983 Report\cite{fofonoff1983algorithms}. The AMOC strength is calculated as the monthly maxima of meridional overturning stream function between 20$^oN$-60$^oN$ and below 450 m depth.
\\
Shortly after the quadrupling of CO$_2$, there is an abrupt transition followed by a dramatic increase in the variance. 
We will apply the indicator to the density difference time series, although one could of course apply the same analysis to the AMOC strength. \\
We choose a window length of 250 data points, corresponding to exactly 20 years of monthly data. Figure \ref{fig:NORESM_4XCO2_Y_colorplot} shows the density difference, $\delta\rho$, color coded according to the values of \textbf{$\Upsilon$}. We only display the part of the time series close to the transition, as this is of primary interest. The point at which the CO2 concentration is abruptly increased, at $t=6000$ months, is indicated by a dashed line.\\ 
The increase in \textbf{$\Upsilon$} during the early part of the AMOC weakening process is apparent. 
Note in particular the three sharp peaks shortly after time t=6000.
Figure \ref{fig:NorESM_4XCO2_qp_plot} again shows the time series, now color coded according to the values of $q$ and $p$. The latter are also plotted for further clarification. From this plot, it becomes clear that the most common fit prior to the transition is the ARMA(1,0) process, which aligns with the observations of \citet{Faranda2015}. After the weakening phase, the value of $p$ is generally an order higher, presumably related to the dramatic increase in the variance. The three sharp peaks in the plot of \textbf{$\Upsilon$} appearing around time $t=6300$ correspond to high values of $q$. The gradual increase in \textbf{$\Upsilon$} preceding these peaks is presumably due to the increase in the persistence (not shown). The $q$ component exhibits peaks prior to $t=6000$, when the forcing is applied and these are reflected in small peaks in $\Upsilon$. These are obviously not connected to the AMOC weakening. Following the initial weakening phase, the value for \textbf{$\Upsilon$} remains high, probably a result of the increase in the $p$ value. 
However, the values of \textbf{$\Upsilon$} do not go above $0.4$ which is considerably smaller than the values found for the 3-box model. In addition, from our previous discussion on the response of the $\Upsilon$ indicator to colored noise, it is conceivable that the increase in $\Upsilon$ observed from in the CESM2 data is primarily caused by changes in the noise amplitude, and not as a consequence of inherent instability of the underlying dynamics.\\
Furthermore we note that, although the result is not explicitly shown, for the CESM2 data $\Delta$BIC$_1$ is always smaller than $\Delta$BIC$_0$, and the $\Delta$BIC$_1$ values are at no point negative, implying that the autoregressive coefficient in the ARMA(1,0) model always satisfy the stationarity constraints. This differs from what was observed in the 3-box model and is presumably related to the difference in the observed $\Upsilon$ values. \\
However, we emphasize that it is not clear if one in actuality can compare values of $\Upsilon$ between datasets. For the autocorrelation and the variance it is typically assumed that it is the change \textit{within} the dataset that is significant, rather than the absolute numerical values.\\
For completeness, we have included a comparison between \textbf{$\Upsilon$} and two other statistical early warning indicators, namely autocorrelation and variance. This is shown in Figure \ref{fig:NorESM_4XCO2_AC_Var_Y_plot}. In all cases, the window length is 250 points, corresponding to approximately 20 years. All three indicators show a clear increase shortly after time t = 6000.  

\subsection{Abrupt $\mathbf{2\times CO_2}$}
\begin{figure}
    \centering
    \includegraphics[width = \linewidth]{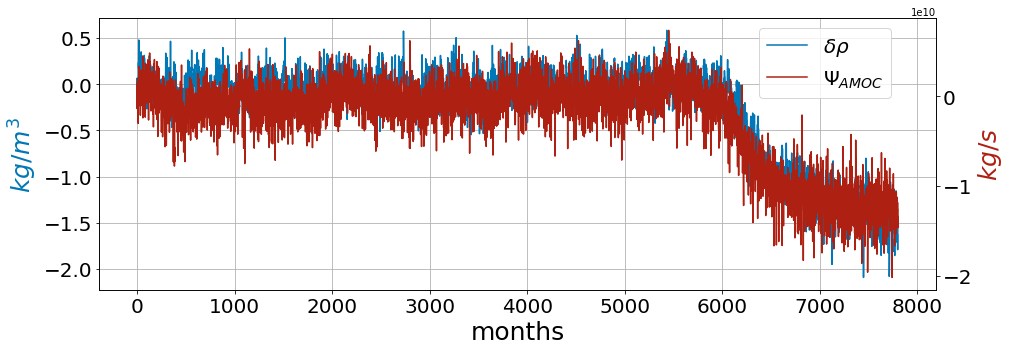}
    \caption{CESM2 model with abrupt $2\times \text{CO}_2$, where the monthly density difference (blue) is plotted together with the maximum AMOC flow strength (red).}
    \label{fig:NorESM_2XCO2_timeseries}
\end{figure}

\begin{figure}
    \centering
    \includegraphics[width = \linewidth]{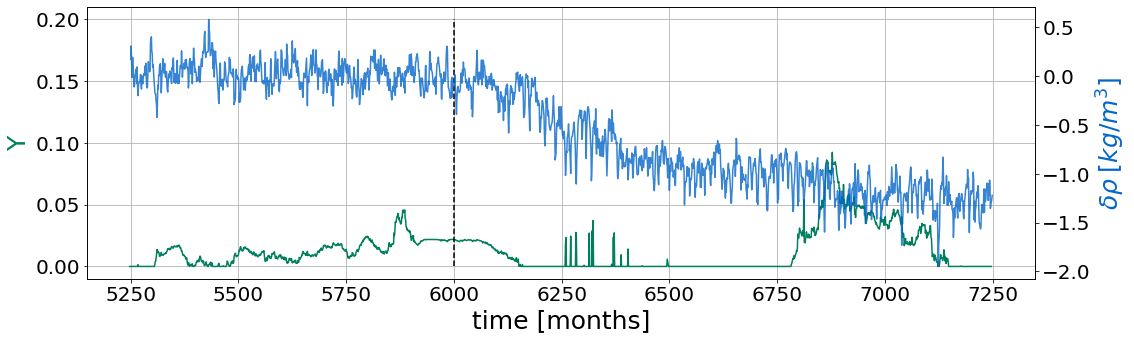}
    \caption{Monthly density changes, $\delta\rho$, for abrupt $2\times \text{CO}_2$ (blue) and the value of \textbf{$\Upsilon$} (green) plotted as functions of time. The dashed line indicates the point when the CO$_2$ concentration abruptly changes.}
    \label{fig:NORESM_2XCO2_Y_colorplot}
\end{figure}

\begin{figure}
\centering
\subfloat(a){%
  \includegraphics[width=\linewidth]{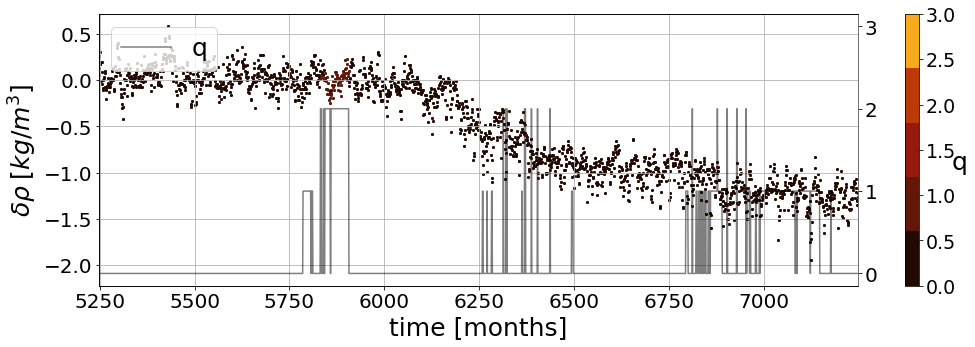}%
  }
\subfloat(a){%
  \includegraphics[width=\linewidth]{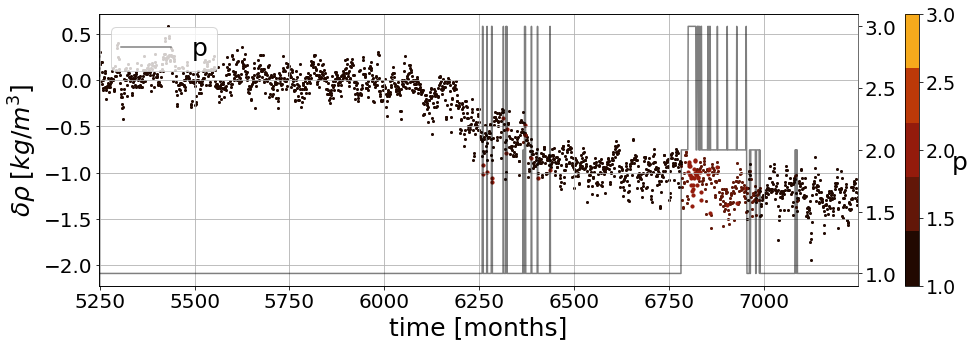}%
  }
\caption{Time series of monthly density changes for abrupt $2\times \text{CO}_2$, color coded according to the value of (a) $q$  and (b) $p$. The value for $q$ and $p$ are also plotted as functions of time in (a) and (b), respectively. }
\label{fig:NorESM_2XCO2_qp_plot}
\end{figure}

\begin{figure}
    \centering
    \includegraphics[width = \linewidth]{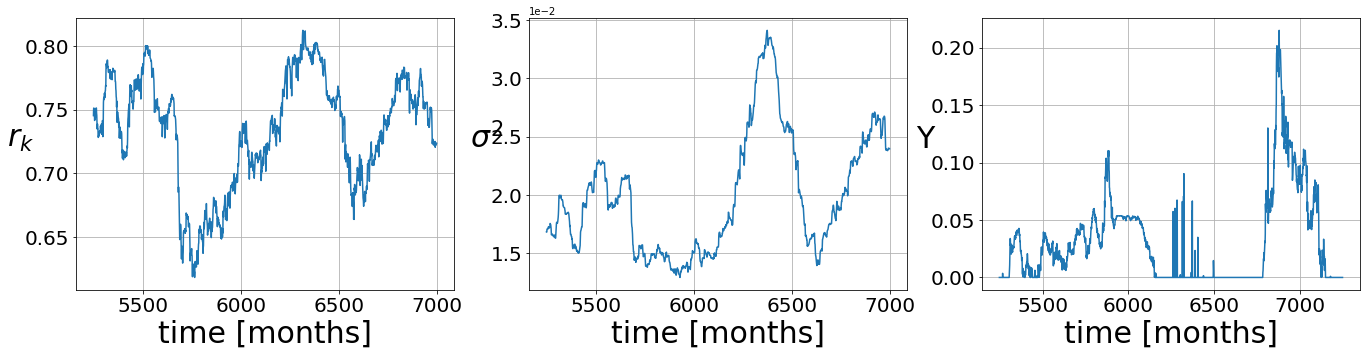}
    \caption{Autocorrelation, variance and \textbf{$\Upsilon$} plotted as functions of time for the case of abrupt $2\times \text{CO}_2$}
    \label{fig:NorESM_2XCO2_AC_Var_Y_plot}
\end{figure}

The time series of the monthly density difference, $\delta\rho$, and AMOC strength, $\psi_{AMOC}$, in the case of abrupt $2\times \text{CO}_2$ is shown in Figure \ref{fig:NorESM_2XCO2_timeseries}. Again, we only apply the indicator to the density difference data, and choose the same window length as in the case of abrupt $4\times \text{CO}_2$. Figure \ref{fig:NORESM_2XCO2_Y_colorplot} shows an excerpt of the density difference time series close to the initial weakening, as well as a plot of the \textbf{$\Upsilon$} values. 
A weakening is clearly seen in the model's own AMOC measure, and is also accurately captured with the measure based on the density difference across the Gulf Stream (Fig. \ref{fig:NorESM_2XCO2_timeseries}).\\ 
The first thing to note is how small the \textbf{$\Upsilon$} values are compared to what we have seen previously; on the order of $10^{-2}$. It should, however, be noted that the $\Delta$BIC values are well above the significance threshold\cite{Preacher2012}. Figure \ref{fig:NorESM_2XCO2_qp_plot} shows the density difference time series color coded according to the value of $q$ and $p$. From this, we again see that prior to the increase in CO$_2$, the most common fit is the ARMA(1,0) process, while after the initial weakening phase the $p$ values show a clear increase. The $q$ value, on the other hand, does not exceed 2, indicating a very low degree of memory in the noise term. Since we have by now clearly demonstrated a correlation with the value of \textbf{$\Upsilon$} and the value of $q$, this should provide an explanation as to why we see such low values of \textbf{$\Upsilon$}. From this analysis, one would conclude the system does not appear to be approaching a tipping point. Indeed, the measure suggests that the weakening in the overturning in this case with reduced forcing is not associated with a loss of dynamical stability. 
Once more we have, as shown in Figure \ref{fig:NorESM_2XCO2_AC_Var_Y_plot}, included a comparison with other early warning indicators. The autocorrelation and variance show a dramatic increase around time t=6000, which corresponds to the appearance of the cluster of sharp peaks in the time series plot for \textbf{$\Upsilon$}.

\section{Discussion}
In summary, we analysed an indicator for dynamical stability based on ARMA modelling as a way to detect transitions in complex systems. A detected need for higher order terms in the ARMA model fitted to moving windows of a timeseries is related to diverging memory properties, which are expected to arise when approaching a transition to a new equilibrium state. The rationale behind this indicator is that it uses a broad family of linear statistical models that can be fitted even on short time series and which have proven their utility in many contexts (see \citet{brockwell_introduction_2002}). That the underlying models do not require long time series is an advantage when employing a sliding window approach on limited data sets.  
The method generalizes classical metrics of instability, and allows one to extract more global dynamical  information from the time series data. \\
The indicator was tested on time series data from a 3-box model of the AMOC, where three categories of critical transitions, namely B-, N-, and R-tipping, were explored. In all cases the transition is identified by the indicator, albeit it is not always easy to interpret the signal.
In the rate-induced tipping scenario a comparison between the avoided tipping and the tipping cases shows a response of the indicator prior to the transition only in the tipping case although the time series are nearly identical at this stage. The indicator also successfully identifies the unstable limit cycle when returning to the upper equilibrium branch. We similarly see fairly clear signals in the bifurcation-induced tipping scenario prior to the transition. For the case of noise-induced tipping, the signal is less clear, obscured by the high amplitude noise. However, when going from the lower to the upper equilibrium branch the indicator signals an increased degree of instability in accordance with the presence of the unstable limit cycle. \\
\\\\
The primary drawback of the \textbf{$\Upsilon$} indicator is that it is computationally quite expensive, at least compared to the autocorrelation and variance, and that, due to its complexity, the results can be harder to interpret. We therefore suggest that the indicator should be applied with care, and preferably in combinations with other measures of instability, like the increase in the order, $p+q$, and the persistence. Although the current scaling with $\tau$, see equation (\ref{eq:Upsilon}), seems to yield reasonable results, it is certainly possible that another scaling would be preferred. It is also possible that this is problem-dependent. This uncertainty regarding the correct scaling is certainly a drawback, but we argue that this problem can largely be circumvented by including an examination of the persistence and order values. However, it would still be advantageous to have an indicator whose values were to have a clear meaning in terms of the stability of the system, and it is not clear if the \textbf{$\Upsilon$} indicator as it stands achieves this, partly due to the aforementioned issue with the choice of the correct scaling. Although we have attempted to make some comparison to other early warning indicators, like the increase in autocorrelation and variance, we are not claiming that the $\Upsilon$ indicator is in any way better than these other indicators, rather that it can act as a complementary approach, as it can allow one to extract more information from time series data. For example, we have suggested, that it might be helpful in identifying the effects of colored noise, something the other indicators struggle with. \\\\
Furthermore, we note that it is conceivable that one would wish to exclude white noise and pure moving-average, MA(1), processes when doing the fitting, as was done in the earlier studies by \citet{Faranda2015}. In such a scenario the modified definition of the $\Upsilon$ indicator would of course no longer be valid, as the ARMA(0,0) process is excluded, and thus cannot be used as a base model. In this case one might argue that the points where $\Delta_1$BIC are negative should either be ignored completely, or one should assume that the best fit is in fact the ARMA(1,0) process and the algorithm is being too strict it its enforcement of the auxiliary conditions on the fitting parameters. This would of course lead to different results than what has been presented here, and is an option worth considering.
\\\\
When considering a full complexity AMOC model as arising from a global climate model (CESM2) many more degrees of freedom are involved. This has two consequences: firstly, the pure categories of tipping cannot really be expected anymore and secondly, the tipping behaviour might disappear altogether as the added degrees of freedom may stabilize the system.  \\
When applied to the CESM2 data, the results were mixed. The measure exhibited a significant increase in \textbf{$\Upsilon$} under the more severe 4xCO$_2$ forcing but much less variability with the weaker 2xCO$_2$ forcing. Hence the measure only registers larger changes in AMOC as associated with dynamically unstable behavior. 
Indeed, it is possible that the model AMOC experiences a continuously shifting steady state, rather than making a transition between two distinct states as in low dimensional models. The results from the doubling CO$_2$ experiment seems to support this hypothesis. Other members of the CMIP6 ensemble exhibiting very different AMOC weakening from the same forcing, with some declining by only 15\% and others falling by 80\% \citep{Madan2022}, and this suggests a continuum of different responses. \\
While the results for $4\times$CO$_2$ suggest a loss of dynamical stability during the AMOC weakening phase, concluding on the tipping behaviour would require a more in depth analysis along the lines done in \citet{hawkins2011}; in this paper the bi-stability is clearly demonstrated by exploring a range of hosing experiments. Although we are confident that the $\Upsilon$ indicator can be used to assess the stability of such complex systems, as was already demonstrated in previous works by \citet{Nevo2017}, concluding on the ability to detect critical transitions would require a full analysis of the hysteresis behaviour of the system.

\begin{table*}
\caption{Adapted from \citet{Alkhayuon2019}}
\begin{ruledtabular}
\begin{tabular}{ccccc}
              &  \textbf{Volume} & \textbf{Salinity} & \textbf{Flux} \\[10 pt]
         \textbf{North Atlantic} & $V_N = 0.3683 \times 10^7$ m$^3$& $S_N = 0.034912$ & $F_N = 0.486$ Sv \\[10pt] 
         \textbf{Tropical Atlantic} &  $V_T = 0.5418\times 10^7$ m$^3$& $S_T = 0.035435$&   $F_T = - 0.997$ Sv \\[10pt] 
         \textbf{Southern Ocean} &  $V_S = 0.6097\times 10^7$ m$^3$& $S_S =  0.034427$ & $F_S = 1.265$ Sv\\[10pt] 
         \textbf{Indo-Pacific} & $V_{IP} =  1.4860\times 10^7$ m$^3$ & $S_{IP} =  0.034668$ & $F_{IP} = - 0.754 $ Sv \\[10pt] 
         \textbf{Bottom Ocean} & $V_B = 9.9250\times 10^7$ m$^3$ & $S_B =  0.034538$ & \\[10pt]
\end{tabular}
\end{ruledtabular}
\end{table*}

\begin{table}
\caption{Adapted from \citet{Alkhayuon2019}}
\begin{ruledtabular}
\begin{tabular}{cccccc}
         \textbf{name} & \textbf{default value} & \textbf{units} & \textbf{name} & \textbf{default value} & \textbf{units} \\[10 pt]
         $\alpha$ & 0.12 & kg/ (m$^{3} \; ^\circ C$) & $K_N$ & 1.762 & Sv \\[10 pt]
         $\beta$ & 790.0 & kg/m$^{3}$  & $K_S$ & 1.872 & Sv\\[10 pt]
         $S_0$ & 0.035 & & $\lambda$ & $1.62 \times 10^7$ & $m^6/(kg\; s)$\\[10 pt]
         $T_S$ & 7.919 & $^\circ C$ & $\gamma$ & 0.36 &\\[10 pt] 
         $T_0 $ & 3.870 &$^\circ C$ & & & \\[10 pt]
\end{tabular}
\end{ruledtabular}
\end{table}

\begin{acknowledgments}
This research has been partly funded by the Deutsche 
Forschungsgemeinschaft (DFG) through grant CRC 1114 „Scaling Cascades in 
Complex Systems“, Project Number 235221301. \\
LaCasce was supported in part by the Rough Ocean 
project, number 302743, from the Norwegian Research Council.
\\
The computations for CESM2 data were performed on resources provided by Sigma2 - the National Infrastructure for High Performance Computing and Data Storage in Norway.
\\
The authors thank Davide Faranda for stimulating discussions, and Amandine Kaiser for help with the development of the code used for the numerical analysis. 
\end{acknowledgments}

\section*{Data Availability Statement}
The data that support the findings of this study are available from the corresponding author upon reasonable request.
\\\\
\nocite{*}
\bibliography{aipsamp}

\end{document}